\documentclass[twocolumn,floatfix]{revtex4-1}
\usepackage{amsmath,amssymb,amsfonts}
\usepackage{amsthm}
\usepackage{graphicx}
\usepackage{color}
\usepackage[utf8]{inputenc}
\usepackage{xspace}
\usepackage[T1]{fontenc}

\newcommand{\ie}{\textit{i.e.}\xspace}
\newcommand{\eg}{\textit{e.g.}\xspace}
\newcommand{\cf}{\textit{cf.}\xspace}
\newcommand{\etc}{\textit{etc.}\xspace}

\newcommand{\rf}{^\star} 
\newcommand{\tind}[1]{^{(#1)}} 
\newcommand{\tinv}[1]{{[0,#1]}} 

\newcommand{\tot}{\ensuremath{_{0}}}
\newcommand{\sys}{\ensuremath{_\text{s}}}
\newcommand{\res}{\ensuremath{_\text{r}}}

\newcommand{\med}{\ensuremath{_\text{med}}}

\newcommand{\rev}{\ensuremath{_\star}}

\newcommand{\conv}{\ensuremath{_\text{aut}}}
\newcommand{\drv}{\ensuremath{_\text{drv}}}

\newcommand{\convrv}{\ensuremath{_{\text{aut},\star}}}
\newcommand{\drvrv}{\ensuremath{_{\text{drv},\star}}}

\newcommand{\can}{\ensuremath{_\text{c}}}
\newcommand{\gc}{\ensuremath{_\text{gc}}}

\newcommand{\den}{\varrho} 

\newcommand{\bent}{\ensuremath{S}} 
\newcommand{\bdkl}{\ensuremath{D}} 
\newcommand{\bz}{\ensuremath{Z}} 

\newcommand{\sent}{\ensuremath{\mathbb{H}}}
\newcommand{\Dklinf}{\ensuremath{\mathbb{\bdkl}}} 

\newcommand{\Z}{\ensuremath{\mathcal{\bz}}} 

\newcommand{\fct}{\ensuremath{O}} 
\newcommand{\Fct}{\ensuremath{\mathcal{O}}} 

\newcommand{\bobs}{\ensuremath{C}} 
\newcommand{\obs}{\ensuremath{\bobs}} 
\newcommand{\Obs}{\ensuremath{\mathcal{\bobs}}} 
\newcommand{\Vobs}{\ensuremath{\vec{\Obs}}} 

\newcommand{\V}{\ensuremath{\mathcal{V}}}
\newcommand{\N}{\ensuremath{\mathcal{N}}}

\newcommand{\R}{\ensuremath{\mathcal{R}}}

\newcommand{\pol}{\ensuremath{\mathcal{M}}}
\newcommand{\field}{\ensuremath{h}}

\newcommand{\fields}{\ensuremath{\vec{\field}}}

\newcommand{\fent}{\ensuremath{\Phi}} 

\newcommand{\Ent}{\ensuremath{\mathcal{\bent}}} 

\newcommand{\bfree}{\ensuremath{F}} 
\newcommand{\Free}{\ensuremath{\mathcal{\bfree}}} 

\renewcommand{\H}{\ensuremath{\mathcal{H}}}
\newcommand{\B}{\ensuremath{\mathcal{B}}}
\newcommand{\U}{\ensuremath{\mathcal{U}}}

\newcommand{\G}{\ensuremath{\mathcal{G}}}

\newcommand{\pars}{ {\ensuremath{\vec{\nu}}} }
\newcommand{\ip}{ {\ensuremath{\vec{\gamma}}} }
\newcommand{\lam}{ {\ensuremath{\vec{\lambda}}} }

\newcommand{\Dkl}{\ensuremath{\mathcal{\bdkl}}} 

\newcommand{\zero}{\texttt{0}\xspace}
\newcommand{\one}{\texttt{1}\xspace}

\newcommand{\df}{\mathrm{\,d}\!\,}
\newcommand{\eav}[1]{\left\langle #1 \right\rangle}

\renewcommand{\L}{\ensuremath{\mathcal{L}}}

\newcommand{\dref}[1]{\ensuremath{\Delta^{\hspace{-.2em}(#1)}_\star\!}} 

\renewcommand{\vec}[1]{\ensuremath{\boldsymbol{#1}}}

\newcommand{\W}{\ensuremath{\mathcal{W}}}
\DeclareMathOperator*{\argmax}{arg\,max}
\DeclareMathOperator*{\cov}{cov}

\begin{document}

\title{Nonequilibrium thermodynamics and information theory:\\
      Basic concepts and relaxing dynamics} 
\author{Bernhard Altaner}
\affiliation{Complex Systems and Statistical Mechanics, Physics and Materials Science Research Unit, University of Luxembourg, Luxembourg}
\begin{abstract}
  Thermodynamics is based on the notions of energy and entropy.
While energy is the elementary quantity governing physical dynamics, entropy is the fundamental concept in information theory.
In this work, starting from first principles, we give a detailed didactic account on the relations between energy and entropy and thus physics and information theory.
We show that thermodynamic process inequalities, like the Second Law, are equivalent to the requirement that an effective description for physical dynamics is strongly relaxing.
From the perspective of information theory, strongly relaxing dynamics govern the irreversible convergence of a statistical ensemble towards the maximally non-commital probability distribution that is compatible with thermodynamic equilibrium parameters. 
In particular, Markov processes that converge to a thermodynamic equilibrium state are strongly relaxing.
Our framework generalizes previous results to arbitrary open and driven systems, yielding novel thermodynamic bounds for idealized and real processes.

\end{abstract}
  \maketitle

\subsection{Introduction}
The basic connections between equilibrium thermodynamics and probability predate modern information theory and go back to Gibbs, Boltzmann and Einstein~\cite{Peliti.Rechtman2017}.
Shannon's information theory has formalized the uncertainty associated to a probabilistic description in an axiomatic way~\cite{Shannon1948}.
The basic ideas of how thermodynamics and information theory are connected have enabled a modern epistemic perspective on thermodynamics~\cite{Jaynes1957,Landauer1961,Procaccia.Levine1976,Bennett1982}.
Extending this perspective to consistently include modern aspects of nonequilibrium thermodynamics is an active field of research \cite{Toyabe_etal2010,Esposito.VandenBroeck2011,Sagawa2012,Berut_etal2012,Horowitz_etal2013,Barato.etal2014,Parrondo.etal2015,Sartori.Pigolotti2015,Rao.Esposito2016}.

The purpose of this article is to describe a universal self-consistent approach to modern nonequilibrium thermodynamics based on the rationale of information theory.
In order to be clear about our scope, we first recall the three elementary basic notions of nonequilibrium processes:
\begin{enumerate}
  \item Irreversible (transient) relaxation of a non-equilibrium system to an equilibrium state specified by thermodynamic parameters.
  \item Driving processes where these thermodynamics parameters are changed.
  \item Non-equilibrium steady states (NESS).
\end{enumerate}
Here, we are concerned with the first two aspects of nonequilibrium thermodynamics.
The third aspect is not covered.
The reason for this is that the thermodynamic consistency of NESS requires a clear notion of \emph{local equilibrium} (and local detailed balance), which builds upon but ultimately goes beyond this work.

Importantly, the relation between non-equilibrium and equilibrium thermodynamics is not that of, for example, the study of elephant and non-elephant metabolism.
The concept of an elephant has nothing to do with the concept of metabolism.
In contrast, thermodynamics always refers back to equilibrium, where there is a clear connection between energy and entropy.

In his autobiographical notes, Einstein expressed his high regard for thermodynamics as a universal theory~\cite{Schilpp.Jehle1951}:
``A theory is the more impressive the greater the simplicity of its premises, the more different kinds of things it relates, and the more extended its area of applicability.
Therefore the deep impression that classical thermodynamics made upon me.
It is the only physical theory of universal content which I am convinced will never be overthrown, within the framework of applicability of its basic concepts.''

Here, we show that even beyond equilibrium, energy and entropy constitute these basic concepts.
In particular, we demonstrate how nonequilibrium thermodynamics emerges as a universal epistemic theory that connects the evolution of energy and entropy to irreversibility and thus to the loss of information in effective physical descriptions.

\subsubsection{Information theory and statistical mechanics}
In 1957, Jaynes published a controversial article entitled ``Information Theory and Statistical Mechanics''~\cite{Jaynes1957}.
In that work, he argues that statistical mechanics can be interpreted as a theory of statistical inference that enables the construction of probabilistic descriptions on the basis of partial knowledge.
He then concludes that the emerging ``subjective'' statistical mechanics should not be regarded as a physical theory.
In particular, according to his view, statistical mechanics does not rely on additional physical dynamical or statistical assumptions.
In order to relate the present work to Jaynes' ideas we briefly revisit his argument now, while a more detailed discussion will follow later.

The connection between a statistical description of a system and a quantitative notion of uncertainty has been formalized by Shannon's pioneering work on information theory~\cite{Shannon1948}:
Given a probability distribution, its \emph{uncertainty} (or \emph{information entropy}) quantifies the expected amount of information acquired when drawing a sample from that distribution.
Operationally speaking, uncertainty  (if expressed in multiples of $1\mathrm{bit} := \ln 2$) is the average number of yes-or-no questions about a system that need to be answered in order to identify the sample unambiguously.

In statistical mechanics, samples correspond to distinct microscopic states of a system.
Usually, the microstate specifies (up to a given resolution) the positions and momenta (and other relevant degrees of freedom) of all entities making up the system.
In contrast, a  macrostate is defined by small number  of macroscopic parameters, like temperature, volume or particle number.
Using a physical model, a macrostate can be inferred from a microstate.
However, the reverse is not true:
In general, there are many microstates that are compatible with a given macrostate.
Consequently, the best we can do is assign a probability distribution to the microstates that is consistent with our macroscopic observations.
Jaynes argues that the least biased, \ie maximally honest guess amongst all these compatible distributions, should maximize its uncertainty.
This so-called maximum entropy (MaxEnt) principle thus constitutes a modern variant of Ockham's razor, which states that the most plausible explanation is the one that requires the least number of additional assumptions.

In his article, Jaynes shows that this maximum entropy distribution coincides with well-known equilibrium distributions, if the statistical averages of macroscopic observables like energy and volume are constrained to observed macroscopic values.
While he does not explicitly motivate his choice of these specific observables, he is well aware that his argument rests on the assumption that these quantities exhibit a very sharp peak.
Ultimately, he motivates his choices \textit{a posteriori} by remarking they are good as long as the resulting statistical theory agrees with experimental behaviour.
He continues showing that the maximum entropy distributions derived by constraining the average energy (and in another example, the volume) correspond to  the well-known equilibrium distributions.
Likewise, the thermodynamic entropy then coincides with the information-theoretic entropy.
He concludes that entropy in the sense of Shannon is the fundamental concept in thermodynamics, even more than the concept of energy itself.
Since entropy does not require any reference to physics, he thus claims that statistical mechanics is simply statistical inference by entropy maximization.
In particular, he argues that his view on statistical mechanics does not require additional physical dynamic and statistical assumptions like ergodicity and equal \textit{a priori} probabilities.

Not surprisingly, Jaynes' approach has attracted a lot of criticism from the physics community.
While the MaxEnt perspective provides an elegant way to circumvent certain difficult questions in statistical mechanics, the ``subjective'' view dismissing the physical nature of the theory is not easily accepted by (statistical) physicists.
At first glance, the subjective nature of entropy seems to be problematic.
After all, the notion of entropy has been used for about a century to construct real machines.
So how can a central thermodynamic quantity depend on the knowledge of the experimenter that measures it?

The answer is that entropy is never measured. 
Even in traditional thermodynamics, the equilibrium entropy of a system is only well defined after one \emph{chooses} the macroscopic parameters to be used in its description.
Or in Jaynes' words~\cite{Jaynes1965}:
``It is clearly meaningless to ask, 'What is the entropy of a crystal?' unless we first specify the set of parameters which define its thermodynamic state.''
As such, entropy is not an \emph{ontological} aspect of a system.
Rather, we should view it as an \emph{epistemic} property that emerges only in the context of a (useful) description. 
Moreover, entropy is never measured in a thermodynamic experiment.
Instead, what one measures in thermodynamics are changes of the energy of a system that come in the form of work and heat.

While this is a strong argument \emph{for} Jaynes' epistemic view of entropy and statistical mechanics, it is similarly a strong argument \emph{against} Jaynes' conclusion that statistical mechanics is not rooted in physics.
The most elementary concept in both thermodynamics and physics in general \emph{is} energy.
Energy and entropy, heat and work are amongst the ``basic concepts'' Einstein spoke about.
Rejecting energy as a fundamental principle of statistical mechanics means disconnecting it from thermodynamics.
Viewing energy as just another extensive quantity whose value is constrained in a MaxEnt principle thus ignores its physical significance.
In particular, energy is different from other extensive quantities.
In thermodynamics, the energy content of a system, usually called \emph{internal energy} is, unlike other extensive quantities like volume, \emph{not directly} accessible.

Moreover, energy is intimately related to physical dynamics.
Thus any approach to thermodynamics that disregards dynamics is at best incomplete and at worst misleading.
After all, in practice we \emph{use} thermodynamics to describe physical \emph{processes}.
In that light, Jaynes perspective on statistical mechanics as a \emph{mere} theory of statistical inference seems indeed questionable.
Moreover, as we will discuss in detail below, also the additional statistical assumption of equal \textit{a priori} probabilities used for the traditional derivation of statistical mechanics is intimately connected to the notion of energy.

In the light of the arguments above there seems only to be one consistent stance on the connection of information theory and statistical mechanics, which will be adopted throughout this work:
We agree with Jaynes' view of statistical mechanics as a \emph{meta}-theory based on statistical inference.
As such, we accept (and indeed hope to promote) the epistemic interpretation of entropy.
However, we firmly disagree with Jaynes' view that statistical mechanics is independent from physics in general and physical dynamics in particular.
In contrast, we will show how the connection between energy, dynamics, statistics and information are at the heart of building a modern self-consistent thermodynamic framework.

\subsubsection{Towards non-equilibrium thermodynamics}
Our goal is to extend and formalize the ideas mentioned above to non-equilibrium, \ie to situations when the state of the system is not appropriately described by an equilibrium (MaxEnt) distribution.
As equilibrium distributions are stationary, a true thermo\emph{dynamics} is necessarily a theory that treats distributions that vary in time.
Temporal changes of the distribution are either due to an autonomous relaxation or due to temporal changes in the thermodynamic parameters that characterize the system and its environment.
The energy of a system may change due to work and heat exchanged with its environment.
The Second Law appears as an inequality relating these quantities.

In order to proceed we thus need a self-consistent approach to the Second Law of Thermodynamics from an information-theoretic perspective rooted in physics:
In particular, our goal is to formalize the epistemic meaning of (irreversible) thermodynamic relaxation as a property of the effective dynamics we use to describe the evolution of a probability distribution.
Intuitively, we treat what we call \emph{strongly relaxing dynamics} which are defined by the property that at any point in time, they discard information about the system's microstate.
More precisely, we consider dynamics that always bring the instantaneous distribution closer to the maximally uncertain one, in accordance with the (instantaneous) thermodynamic control parameters.
Mathematically, this property comes in the form of an inequality for the evolution of an information-theoretic measure for the difference of a (nonequilibrium) distribution from the corresponding \emph{reference equilibrium state}~\cite{Shore.Johnson1980}.
After all, relaxation is intimately related with the notion of irreversible processes, \ie changes in the state of the system that discard information about the system's past. 

From this single information-theoretic inequality, we will be able to derive the Second Law of Thermodynamics and other related inequalities, like the law of exergy degradation, a general version of Landauer's principle and other work inequalities.
In fact, we show that these inequalities are all equivalent.
While some of the results that we present here have been derived previously in more specialized contexts~\cite{Procaccia.Levine1976,Landauer1961,Jarzynski1997,Vaikuntanathan.Jarzynski2009,Takara.etal2010,Esposito.VandenBroeck2011}, we believe that our presentation comprises the most general consistent formulation of nonequilibrium thermodynamics so far.
In particular, it goes beyond the canonical or grandcanonical ensemble to arbitrary situations where a system can exchange a conserved extensive quantity (like particles, volume, energy) with its environment.
Moreover, our formulation applies equally to situations where a system is \emph{driven} (\ie where the control parameters of the system are changed with time) and to systems that reside in \emph{changing environments}.

In particular, our theory predicts the (non-integrable) \emph{non-isothermal} corrections that appear if the temperature of the environment is changed during a non-isothermal process.
It further emphasizes the special role played by (inverse) temperature if compared to other intensive environment parameters (like chemical potentials or pressure).
As a consequence, the special role and general importance of the free energy (which appears as a generating function in the canonical ensemble) becomes evident.
Further, we introduce equilibrium and nonequilibrium versions of a ``natural'' thermodynamic potentials, a general notion for the ``heat content'' of a system  and a notion of ``exergy'' referring to the potential of a system to perform work in an autonomous relaxation process.

A crucial point to keep in mind during the rest of this work is the distinction between thermodynamic quantities that are \emph{controlled} by an experimenter, and thermodynamic quantities that are \emph{measured}, and thus show an autonomous dynamics prescribed by the system relaxation dynamics.
This leads to a new separation of the power (\ie the rate of work) into a ``driven'' and a ``autonomous'' part which both can have either sign.
Although this distinction is related to the distinction between ``input'' and ``output'' power in macroscopic heat engines, it is conceptually different.
We will further see that the \emph{equalities} that hold for reversible processes have analogous \emph{inequalities} for real (irreversible) processes.
As the present framework is built on information-theoretic ideas, we also hope to convince the sceptical reader that the conversion of information to work (and \textit{vice versa}) is a very real and fundamental property of any thermodynamic description.

After having provided the basic intuitive concepts of our work, the main part of this text is dedicated to a formal, yet didactic mathematical treatment of these ideas.
Nothing more complicated than (multidimensional) real analysis is required.
As such, the present work aims to be self-contained.
We hope it is comprehensible for readers with an undergraduate course in physics or mathematics.

Our work is structured as follows:
In section \ref{sec:canonical-ensemble} we quickly review the basic ideas of Gibbs that lead to the canonical ensemble for situations where a system exchanges energy with its environment~\cite{Gibbs1902}.
Following the derivation of the grandcanonical ensemble for open systems that can exchange matter with their environment, we formulate equilibrium ensembles for general open systems in section~\ref{sec:generalized-ensembles}.
The formal notions of information theory are introduced in section~\ref{sec:information-theory}.
With these prerequisites, in section~\ref{sec:relaxing-dynamics} we formalize the relaxation condition and see how it naturally appears for Markovian, \ie memoryless, dynamics.
Section~\ref{sec:noneq-td} defines non-equilibrium state variables and process quantities like work and heat.
Using these definitions, we show the equivalence of the information-theoretic relaxation condition with the Second Law, where the latter appears as the statement that the irreversible entropy production is never negative.
In section~\ref{sec:derived-concepts} we treat further thermodynamic notions like that of heat content and derive (in-)equalities for (ir-)reversible processes that relate (ir-)reversible work and information.
We discuss our results in a broader context and  conclude in section~\ref{sec:discussion}.

\subsection{Canonical statistical mechanics}
\label{sec:canonical-ensemble}
In this section we briefly revisit the historic approaches to equilibrium statistical thermodynamics, which date back to Gibbs and Boltzmann.
We will not dive into the technical details, but rather use this quick review in order to introduce our notation.
For more details, we refer to standard textbooks like Refs.~\cite{Callen1998,Penrose1970}.
However, in order to understand the notion of generalized ensembles in the next section requires a basic intuition about the derivation of the canonical ensemble, \ie of systems that exchange energy with their environment.

The historic and still most common approach towards a statistical formulation of thermodynamics starts from Hamiltonian mechanics as the fundamental \emph{reversible} dynamics on the lowest level.
A system with microscopic states $\vec{z}$ is characterized by the Hamiltonian $H\tot(\vec{z})$, which specifies the energy of a physical configuration.
Since energy is a conserved quantity, the phase space volume $\W(\U)$ of all states with a fixed energy $H\tot(\vec{z}) = \U$ is a measure for the number of (physically distinguishable) states that are accessible to the system.
One proceeds by defining the function $\Ent(\U):= \ln \W(\U)$ and its derivative $\beta := \frac{\df \Ent(\U)}{\df\U}$.
Note that throughout this work we set Boltzmann's constant $k_\text{B}\equiv 1$ to unity.

In his seminal work on statistical mechanics~\cite{Gibbs1902}, Gibbs showed that $\beta^{-1}$, $\Ent$ and $\U$ fulfil the phenomenological relations for temperature, entropy and internal energy that were already known from phenomenological (equilibrium) thermodynamics.
Today, we call this the microcanonical derivation and understand it as a first-principles approach to the thermodynamics of isolated systems.%

A useful application of thermodynamics requires us to go beyond isolated systems and treat systems in contact with an environment that acts as an energy reservoir.
To that end, we split a large isolated system with degrees of freedom $\vec{z} = (\vec{x},\vec{y})$ into a system with degrees of freedom $\vec{x}$ and environment degrees of freedom $\vec{y}$.
In general, the Hamiltonian $H\tot(\vec{z}) = H\sys(\vec{x}) + H\res(\vec{y}) + H_\text{i}(\vec{x},\vec{y})$ of the large system is split into the energy function of the system and the environment, $H\sys(\vec{x})$ and $H\res(\vec{y})$, and an interaction term $H_\text{i}(\vec{x},\vec{y})$.

One is interested in the marginal distribution $\den(\vec{x}) \propto \int \den(\vec{x},\vec{y}) \df \vec {y}$ of the system degrees of freedom.
Without going into the details of the derivation, the typical approach is to start with the microcanonical (\ie fixed-energy) ensemble of the large combined system at energy $\U\tot$.
Because $\den(\vec{x})$ is proportional to the number of possible configurations of the environment for a given value of the system energy $E(\vec{x})$, it follows that the marginal distribution for the system degrees of freedom reads
\begin{align}
  \den (\vec{x}) = \frac{1}{\Z\can}\exp\left( -\beta E(\vec{x}) \right).
  \label{eq:canonical-density}
\end{align}
where the canonical partition function
\begin{align}
  \Z\can = \int\exp\left( -\beta E(\vec{x}) \right)\df \vec{x}
  \label{eq:canonical-partition-function}
\end{align}
ensures normalization.
Usually, the energy function of the system is simply the system part of the Hamiltonian, $E(\vec{x})=H\sys(\vec{x})$.
More precisely, the argument uses the extensivity of the energy (\ie its scaling in the thermodynamic limit) and requires the assumption of short-range interactions, meaning that in this limit the energetic contribution $H_\text{i}(\vec{x},\vec{y})$  due to system-bath interactions is negligible compared to both $H\sys(\vec{x})$ and $H\res(\vec{y})$.
Moreover, and crucially, the argument (often done using a Taylor expansion of the bath entropy $\Ent\res$ around the expectation value of the bath energy), uses the fact that energy is conserved.
Similar to the microcanonical case, the factor $\beta :=\partial_{\U\tot} S\tot $ relates changes of phase space volume to energy changes in the isolated combined system and is thus interpreted as its inverse temperature.
Under the assumption that the environment is much larger than the system, $\beta$ is determined by the inverse temperature of the heat bath alone.
From the system's perspective, $\beta$ fixes the expected value of the system energy
\begin{align}
  \U  := \eav{E} ,
  \label{eq:system-internal-energy}
\end{align}
where the equilibrium (or reference) average of a phase-space function $\fct(\vec{x})$ is defined as
\begin{align}
  \eav{\fct} :=\int\den (\vec{x})\fct(\vec{x}) \df \vec{x}.
  \label{eq:average-ref}
\end{align}
Introducing the entropy functional $\sent$ of a probability density $\den(\vec{x})$ as
\begin{align}
  \sent[\den] := -\int \den(\vec{x}) \ln \den(\vec{x}) \df \vec{x},
  \label{eq:shannon-entropy}
\end{align}
the equilibrium (or Gibbsian) entropy of the system reads
\begin{align}
  \Ent  := \sent[\den ].
  \label{eq:Gibbs-ent}
\end{align}
Mathematically, the negative logarithm of the canonical partition function $\Z\can$
\begin{align}
  \fent\can  := - \ln \Z\can,
  \label{eq:dimless-potential}
\end{align}
is a cumulant-generating function~\cite{Touchette2009}.
We will thus refer to it as the natural dimensionless or generating ``potential''.
Thermodynamically, it appears as a dimensionless potential defined by the Legendre duality relation
\begin{align}
  \Ent  + \fent\can  = \beta\U.
  \label{eq:canonical-duality}
\end{align}
Multiplying this dimensionless potential by the temperature $\beta^{-1}$, we obtain the free energy
\begin{align}
  \Free  := \U  -\beta^{-1}\Ent  \equiv \beta^{-1}\fent\can 
  \label{eq:canonical-free-energy}
\end{align}
as the natural dimensional potential for the canonical situation.

Notice that we use the first equality to \emph{define} the free energy.
This definition will remain valid and important in the rest of this work, even in the context of open systems, where the natural generating potential is \emph{not} obtained from a \emph{canonical} partition function.

\subsection{Open systems and generalized ensembles}
\label{sec:generalized-ensembles}

In general, the energy function of the system $E(\vec{x})$ depends on a set of parameters.
On the one hand, there are given unchangeable microscopic parameters like fundamental physical constants. 
On the other hand, there are thermodynamic system parameters like its volume $V$, the numbers $\vec{N} = (N_k)_k$ of particles of different species $k$,  external fields $\fields = (\field_l)_l$ \etc.
In the following, we will summarize these macroscopic \emph{control parameters} by the symbol $\lam$ and occasionally make the dependence of the system energy function explicit by writing $E = E(\vec{x};\lam)$.

Amongst the system control parameters, a special role is taken by extensive quantities (like volume and matter).
While in a closed system, the value of these quantities is fixed, a corresponding open system can spontaneously exchange such extensive quantities with its environment.
Consequently, an extensive system control parameter for a closed system becomes a \emph{measurable observable} in the corresponding open system.
In particular, we are interested in the exchange of conserved extensive quantities, that cannot be destroyed or created by neither the system nor the environment.
Then, any change of these quantities in the system must come from an exchange of the quantity with the system's environment acting as a corresponding thermodynamic reservoir.
As such, the present treatment excludes systems that allow for chemical reactions, where one species of matter may be transformed into another one.

While this situation has a strong analogy with the transition from the microcanonical to the canonical ensemble, notice that there is a crucial difference:
The energy function $E$ (or more precisely, its average $\U$), is not measurable, while the quantities associated to macroscopic extensive control parameters are.
Moreover, the energy is not just another phase space function, but, as we have mentioned in section~\ref{sec:canonical-ensemble}, a quantity that is intimately related to the fundamental microscopic reversible \emph{dynamics} and thus the microcanonical definition of entropy itself.
Keeping this special role of the energy in mind, we may not be surprised by the emergence of peculiar terms for systems in contact with environments that undergo temperature changes.
Are more detailed discussion if the special role of energy is postponed to section~\ref{sec:discussion}.

For concreteness, we briefly review the case of the grandcanonical ensemble for a system that can exchange both energy and particles with its environment.
For simplicity, we assume that there is only a single species of particles.
In the grandcanonical ensemble, a system microstate $\vec{x}$ is an element of an extended phase-space, sometimes called Fock space.
The Fock space contains microscopic configurations for any number of particles, and the system may switch between different ``particle shells''.
As said above, the number of particles $N$ then becomes an observable $N(\vec{x})$ and thus a function of the microstate, instead of a system parameter.

The number of particles is an extensive quantity and thus its asymptotic statistics behave regularly in an appropriate thermodynamic limit.
Accordingly, there is a corresponding intensive  parameter called the chemical potential $\mu$.
Assuming the environment to be much larger than the system, \ie pretending that it behaves as an infinitely large particle reservoir, we characterize it solely by the value of $\mu$.
Because the number of particles is conserved, the chemical potential characterizing the equilibrium configuration for the system degrees of freedom needs to be the same in the system and its environment.
Marginalization to the system degrees of freedom thus yields the grandcanonical distribution function
\begin{align}
  \den (\vec{x}) := \frac{1}{\Z\gc}\exp\left( -\beta(E(\vec{x};\lam) -\mu N(\vec{x}))  \right).
  \label{eq:grand-canonical-density}
\end{align}
with the grandcanonical partition function
\begin{align}
  \Z\gc := \int \exp\left( -\beta(E(\vec{x};\lam) - \mu N(\vec{x}))  \right) \df{\vec{x}}.
  \label{eq:grandcanonical-partition-function}
\end{align}
The equilibrium entropy is again given by the entropy functional \eqref{eq:Gibbs-ent}, where this time we insert the grandcanonical distribution, Eq.~\eqref{eq:grand-canonical-density}.
For any particular value of $\mu$ and $\beta$ it determines the equilibrium expectation value of the particle number
\begin{align}
  \N := \eav{N}
  \label{eq:grand-canonical-particle-number}.
\end{align}
The grandcanonical dimensionless generating potential $\fent\gc := - \ln \Z\gc$ obeys the Legendre duality relation
\begin{align}
  \Ent  + \fent\gc  = \beta(\U  - \mu\N ).
  \label{eq:grand-canonical-duality}
\end{align}
Its dimensional version is obtained by multiplication with temperature $\beta^{-1}$ and is called the grand potential
\begin{align}
  \Omega &:=  \U  - \mu\N  - \beta^{-1}\Ent  \label{eq:grand-potential}\\
  &= \beta^{-1}\fent\gc.  \nonumber
\end{align}

The same logic used in deriving the grandcanonical ensemble can also be applied by turning the volume $V$ from a control parameter to an observable $V(\vec{x})$.
Then, the corresponding intensive quantity is the (negative) pressure, which specifies the equilibrium expectation value $\V = \eav{V}$ in the so-called isobaric ensemble.
For generality and ease of notation, we will work with a general ensemble where some control parameters describing conserved extensive quantities have been transformed into extensive observables.
To that end, we consider a system in a heat bath at temperature $\beta$.
As before, \emph{system control parameters} are collectively denoted by $\lam$.
All uncontrolled extensive quantities (other than the energy $E(\vec{x})$ with $\eav{E}=\U$) have associated observables $\vec{\obs}(\vec{x})$, with corresponding averages
\begin{align}
  \Vobs := \eav{\vec{\obs}}.
  \label{eq:observable-averages}
\end{align}
The conjugate intensive environment parameters are denoted by $\ip$.
The generalized distribution thus reads 
\begin{align}
  \den _\pars = \frac{1}{\Z}\exp(-\beta(E(\vec{x};\lam) - \ip \vec{\obs}(\vec{x}))),
  \label{eq:general-maxEnt-distribution}
\end{align}
where the vector $\pars=(\beta,\ip,\lam)$ summarizes all thermodynamic parameters.
In the grandcanonical example, the system parameters are the volume $V$ and fields $\fields$.
Hence, we have $\lam = (V,\fields)$ and $\left(\ip,\vec{\obs}(\vec{x})\right) = \left(\mu,N(\vec{x})\right)$.

Again, in the thermodynamic limit, the appropriate entropy is given by Eq.~\eqref{eq:Gibbs-ent} and obeys the general Legendre duality
\begin{align}
  \Ent  + \fent  = \beta(\U  - \ip \Vobs ).
  \label{eq:general-Legendre-duality}
\end{align}
for the generating potential
\begin{align}
  \fent  := -\ln\Z
  \label{eq:general-dimless-potential}.
\end{align}
Previously, variants of this potential have been called (in analogy to the free energy) the (negative) \emph{free entropy} or the Massieu(--Planck) potential, \cf~Refs.~\cite{Callen1998,Balian2007,Touchette2009}.
However, in order to avoid confusion in situations where these nomenclature requires the definition of a \emph{specific} physical situation, we will just call it the ``natural'' or ``generating'' adimensional potential.
Its dimensional variant 
\begin{align}
  \G  := \beta^{-1}\fent  \equiv \U  - \ip \Vobs  - \beta^{-1}\Ent 
  \label{eq:general-dimensional-potential}
\end{align}
is called the natural energetic potential or \emph{natural energy}.
It can be understood as the generalization of the grand potential, Eq.~\eqref{eq:grand-potential}, to generalized open systems.
Notice that even though structurally similar, the free energy $\Free := \U - \beta^{-1}\Ent$ is a distinct well-defined quantity for any generalized ensemble.
As we will see below, it is crucially related to the notion of work in isothermal processes.
In contrast, the definition of the natural dimensional potential $\G$ depends on the ensemble, \ie on the question which extensive macroscopic parameters are controlled and which are measured.
Hence, the natural energy $\G$ must \emph{not} be understood as a mere generalization of the free energy $\Free$.

The intensive bath parameters $\ip$ are (energetically) conjugate to the extensive quantities that we obtained by changing from system control parameters to system observables.
In addition there are energetically conjugate quantities for the remaining control parameters, defined as 
\begin{align}
  \R_i = \eav{\frac{\partial E}{\partial \lambda_i}} = \beta^{-1}\frac{\partial \fent}{\partial \lambda_i},
  \label{eq:system-conjugate-qantities}
\end{align}
where $i$ denotes the $i$th component of $\lam$.
For the grandcanonical ensemble with $E=E(\vec{x};V,\field)$, they are the (negative) internal system pressure
\begin{align}
  -p := \eav{\frac{\partial E}{\partial V}} = T\frac{\partial \fent}{\partial V}
  \label{eq:pressure}
\end{align}
and the polarisations $\vec{\pol}$
\begin{align}
  \vec{\pol} := \eav{\frac{\partial E}{\partial \fields}}=T\frac{\partial \fent}{\partial \fields}
.
  \label{eq:polarisation}
\end{align}
Notice that Eq.~\eqref{eq:system-conjugate-qantities} is one of the parameter-derivatives of the generalized potential:
\begin{subequations}
\begin{align}
  \frac{\partial\fent}{\partial \beta} &= \U-\ip\Vobs,\\
  \frac{\partial\fent}{\partial \lambda_i} &= \beta \R_i,\\
  \frac{\partial\fent}{\partial \gamma_i} &= -\beta \Obs_i. 
\end{align}
\label{eq:parameter-derivatives-generating}%
\end{subequations}%
Assuming analyticity of $\fent$ in the parameters $\pars$ we further obtain the Maxwell relations:
\begin{subequations}
  \begin{align}
    \frac{\partial \R_i}{\partial \lambda_j}&=\frac{\partial \R_j}{\partial \lambda_i}\\
    \frac{\partial \Obs_i}{\partial \gamma_j}&=\frac{\partial \Obs_j}{\partial \gamma_i}\\
    \frac{\partial \R_i}{\partial \gamma_j}&=-\frac{\partial \Obs_j}{\partial \lambda_i}\\
    \R_i + \beta \frac{\partial \R_i}{\partial \beta} &= \frac{\partial \U}{\partial \lambda_i}-\ip\frac{\partial \Vobs}{\partial \lambda_i}\\
    \beta \frac{\partial \Obs_i}{\partial \beta} &= -\frac{\partial \U}{\partial \gamma_i}+\ip\frac{\partial \Vobs}{\partial \gamma_i}
  \end{align}
  \label{eq:Maxwell-relations}%
\end{subequations}%
While the above relations are universal and follow directly from the elementary theory, the dependence of equilibrium state variables on parameters is specific to the system and its environment.
In traditional thermodynamics the algebraic relations between parameters and observables are called equations of state.
They are uniquely determined by $\fent$ as a function of $\pars$.

\subsection{Information theory and thermodynamics}
\label{sec:information-theory}

The canonical treatment of systems that can exchange energy and other extensive quantities with a large equilibrated environment is usually considered as a derivation of thermodynamics from first principles.
But what are these first principles exactly?

While Jaynes argued that the nature of the dynamics (and thus questions of ergodicity) do not matter for the notion of thermodynamics as a physical (meta-)theory, we disagree with this statement.
In contrast, we argue that for a consistent thermodynamics on some effective level, we always need to refer back to an appropriate \emph{microscopic dynamics}, which is \emph{reversible} and has a fundamental constant of motion (which we call energy), which \emph{generates} the dynamics.
From that perspective, it does not matter whether we choose quantum mechanics or classical Hamiltonian mechanics as the corresponding underlying reversible theory.

While the fundamental requirement on a microscopic dynamics connects energy and reversibility, we need a second requirement that introduces probability into the description.
In the canonical derivation, this is achieved by the assumption of equal \textit{a priori} probabilities for isolated systems, where the dynamics is completely specified by the energy function attributed to \emph{all} degrees of freedom.
As such, it amounts to the choice of a ``fair'' reference measure for the microstates of a microscopic dynamics.
We will discuss this perspective in more detail in Sec.~\ref{sec:discussion}.

In the derivation of the canonical (generalized) ensembles for (open) systems, the statistical quantity that is consistent with the concept of thermodynamic entropy has emerged as the universal functional $\Ent  =\sent[\den ]$ of the corresponding equilibrium distribution $\den$.
The independence of this observation from the details of the physical situation has its origin in the exponential scaling of the marginal probability density functions for \emph{extensive} quantities in the thermodynamic limit.
It has recently been understood and mathematically formalized using the theory of large deviations~\cite{Ellis2012,Touchette2009,Chetrite.Touchette2014}.
Similarly, the physical interpretation of the parameters $\beta$ and $\ip$ as intensive parameters (temperature, chemical potentials, pressure) requires the thermodynamic limit for the corresponding reservoirs of extensive quantities (energy, particles, volume).

As we explained in some detail in the introduction, the epistemic approach is rooted in Shannon's information theory~\cite{Shannon1948,Shore.Johnson1980,Cover.Thomas2006}.
In information theory, the entropy functional $\sent[\den]$ is an axiomatically defined quantitative measure of the \emph{uncertainty} of a probability distribution.
Or differently interpreted, the entropy of a probability distribution represents the average \emph{information} one obtains from sampling a state $\vec{x}$ according to $\den(\vec{x})$.
In the epistemic perspective, equilibrium statistical mechanics is a theory of statistical inference, which assigns a probability to a microstate given the prior knowledge of a microscopic physical model (specified by an energy function $E(\vec{x})$ and other observables), macroscopic parameters and macroscopic observations.
The equilibrium reference distribution $\den$ is obtained as the maximally non-committal (\ie the least biased) distribution that is consistent with given knowledge about the expectation values of the global extensive quantities.
Consequently, this distribution should have the maximal uncertainty amongst all distributions  that yield the correct macroscopic averages.

Formally, amongst all $\den'$ that obey the constraints $\eav{E}_{\den'} = \U$ and $\eav{C}_{\den'} = \Obs$, the maximally uncertain equilibrium distribution $\den $ is \emph{defined} as
\begin{align}
  \den  := \argmax_{\den'} \sent[\den'].
  \label{eq:maxEnt}
\end{align}
In this so-called MaxEnt approach, the intensive environment parameters $\beta$ and $\ip$ enter as the Lagrange multipliers in the maximization problem.
Further, the partition function $-\ln(\Z)$ appears as the Lagrange multiplier that ensures normalization.

The main difficulty with the generalization of this idea to non-equilibrium situations is evident:
in equilibrium, the macroscopic state of a system is characterized by a set of independent thermodynamic state variables, which fix the reference expectation values.
Equilibrium equations of state then unambiguously give the values of the remaining thermodynamic variables.
Out of equilibrium, there is no general notion of the ``correct'' macroscopic expectation values or their mutual relations.

Information theory yet provides us with a means to quantify the disparity of a distribution $\den$ from some reference distribution $\den\rf$ by means of the Kullback--Leibler (KL) divergence~\cite{Shore.Johnson1980,Cover.Thomas2006}:
\begin{align}
  \Dklinf[\den\|\den\rf] := \int \den(\vec{x}) \ln \frac{\den(\vec{x})}{\den\rf(\vec{x})} \df \vec{x} \geq 0
  \label{eq:dkl}
\end{align}
Non-negativity of this quantity follows from Jensen's inequality, and (ignoring the more subtle notions of measure theory), $\Dklinf[\den\|\den\rf]=0$ holds if and only if $\den=\den\rf$.

The KL-divergence has a natural interpretation in terms of coding theory, where we think of events (or in the present context, microstates) as encoded by symbolic sequences.
In binary coding, such a symbol sequence can be thought of as the answers to a specific set of yes-or-no questions about the microstate.
With that idea in mind, the Shannon--Kraft--McMillan coding theorem states that for any probability distribution $\den$ there exists an optimal natural ``encoding'' (or, equivalently, an optimal set of questions), such that the expected sequence length is proportional to $\sent[\den]$~\cite{Cover.Thomas2006}.
The KL-divergence quantifies the additional average message length (\ie the additional average number of questions that need to be answered) when one draws samples from a distribution $\den$, but uses an encoding that was optimized for another distribution $\den\rf$~\cite{Shore.Johnson1980}.

With this interpretation it is not surprising that the KL-divergence will play an important role for formulating nonequilibrium thermodynamics:
after all, we intend to make meaningful statements about nonequilibrium states (\ie, nonequilibrium probability distributions) by referring to the statistical notions that arise in the context of equilibrium systems (with their corresponding equilibrium distributions).

\subsection{Relaxing dynamics}
\label{sec:relaxing-dynamics}
A crucial property of an equilibrium distribution is its stationarity.
Hence, a true thermo\textit{dynamics} is necessarily a nonequilibrium theory that deals with time-dependent processes.
As said above, here we consider two aspects of thermodynamic processes.
\emph{Relaxation processes} govern the convergence of a non-equilibrium distribution to its equilibrium state.
The latter is defined by the values of the thermodynamic parameters $\pars=(\beta,\ip,\lam)$ that characterize the system and its environment.
In addition, an external agent may \emph{drive} the system by applying a thermodynamic protocol $\pars = \pars(t)$ that specifies how the thermodynamic parameters change with time.

In what follows, we characterize the instantaneous (non-equilibrium) state of the system at time $\tau$ by a time-dependent probability distribution $\den\tind{\tau}$.
Its evolution to a later state is governed by the first-order (partial) differential equation
\begin{align}
  \frac{\partial}{\partial\tau}{\den}\tind{\tau} = \L_{\pars(\tau)}[\den\tind{\tau}],
  \label{eq:evolution-operator}
\end{align}
where the dynamics is encoded in the operator $\L_\pars$.
At this point, we demand nothing of the operator besides that it maps a probability densities to probability densities and that its value $\L_{\pars(\tau)}[\den\tind{\tau}]$ is a state function, \ie, that it depends only on the instantaneous $\den\tind{\tau}$.
If the dynamics is a model for the evolution of the state of a thermodynamic system, the operator necessarily depends on the value of the physical control parameters $\pars$.

A thermodynamic system out of equilibrium that is left on its own in a fixed environment will eventually equilibrate in what we call the relaxation process.
We say that a dynamics is \emph{weakly relaxing} if for any constant value of the physical parameters $\pars$ we have 
\begin{align}
  \den\tind{\tau} \xrightarrow{\tau\to\infty}  \den\rf_\pars.
  \label{eq:relaxation-condition}
\end{align}
Naturally, the asymptotic distribution  $\den\rf_\pars$ is the stationary equilibrium distribution for the appropriate physical situation, which is defined by the right-hand side of Eq.~\eqref{eq:general-maxEnt-distribution}.
In what follows we emphasize the role $\den\rf_\pars$ as a thermodynamic reference distribution (for given parameter values $\pars$) by using the $\star$-symbol as a superscript.

In the last section, we introduced the Kullback--Leibler divergence $\Dkl[\den\Vert\den\rf]$ as the information-theoretic distance of a distribution $\den$ from a reference distribution $\den\rf$.
Thus, it is natural to define the thermodynamic distance
\begin{align}
  \Dkl\tind{\tau}_{\pars}:=
  \Dklinf\left[ \den\tind{\tau}\middle\|\den\rf_{\pars} \right].
  \label{eq:dklref}
\end{align}
It measures the disparity of the instantaneous non-equilibrium distribution $\den\tind{\tau}$, which evolves according to $\L_\pars$ from the corresponding equilibrium reference distribution $\den\rf_\pars$.
In the context of the present interpretation is has first appeared in Ref.~\cite{Procaccia.Levine1976} under the name of ``entropy deficiency'', but constitutes also as a central quantity in more recent work ~\cite{Vaikuntanathan.Jarzynski2009,Takara.etal2010,Esposito.VandenBroeck2011,Esposito.VandenBroeck2011}.

With the definition of the thermodynamic distance, we reformulate the condition of a weakly relaxing dynamics at constant parameters $\pars$ as
\begin{align}
  \Dkl\tind{\tau}_{\pars} \xrightarrow{\tau\to\infty}0.
  \label{eq:relaxing-DKL}
\end{align}
Weakly relaxing dynamics are thus only constrained by the their time-\emph{asymptotic} behaviour.
For driven systems, where parameters $\pars$ may change with time, the weak relaxation condition is not appropriate.
Even for non-driven systems that only relax, a meaningful thermodynamic for transient behaviour should require a constraint formulated on the dynamics \emph{at all times}, rather than only referring to its final destination.

The central property of thermodynamic relaxation is its \emph{irreversibility}:
it is impossible to revert the effects of relaxation without spending something a form of energy that thermodynamics calls work.
However, as the goal of the epistemic approach is to understand thermodynamic concepts from probabilistic and information-theoretic concepts, and not the other way around, we will not introduce work yet.
Instead we propose more abstractly that \emph{irreversibility} simply encodes the \emph{loss of information}, \ie the \emph{increase of uncertainty}.

We already have everything in place to define what we call a \emph{strongly relaxing dynamics}:
A dynamics~\eqref{eq:evolution-operator} is \emph{strongly relaxing} if and only if at any moment in time and for any probability distribution, we have
\begin{align}
  \frac{\partial}{\partial \tau} \Dkl\tind{\tau}_{\pars} \leq 0,
  \label{eq:strongly-relaxing}
\end{align}
with equality if and only if $\den\tind{\tau} = \den\rf_\pars$, and thus $\Dkl\tind{\tau}_\pars = 0$.
This condition is equally meaningful for driven and non-driven situations.
We emphasize that on the left-hand side of this condition, a \emph{partial} time-derivative appears.
As such, the time-derivative acts only on $\den\tind{\tau}$, which is in turn calculated using the evolution rule Eq.~\eqref{eq:evolution-operator}.
Practically it means that the evolution operator $\L_\pars$ appearing in Eq.~\eqref{eq:evolution-operator}, as well as the reference distribution $\den\rf_\pars$, are evaluated at their instantaneous values $\pars = \pars(\tau)$.
For driven systems, a total time-derivative  would includes an additional term proportional to $\dot{\pars}(\tau)$ multiplied by a parameter gradient.
We will encounter and analyse such parameter gradients in more detail below.
For now let us just mention that for constant parameters strong relaxation implies weak relaxation. 

Let us briefly stop here to emphasize the following:
The strong relaxation condition formulated as the inequality~\eqref{eq:strongly-relaxing} is the single dynamical assumption we will use in the remainder of this work.
In particular, it is the single condition that we will need to \emph{derive} all the well-known thermodynamic inequalities like the Second Law.
As we will see below, all these thermodynamic process inequalities are in fact equivalent.

The crucial point about the inequality~\eqref{eq:strongly-relaxing} is that it has an unambiguous information-theoretic meaning:
a strongly relaxing dynamics constantly discards information.
Or more mathematically:
The distribution $\den\tind{\tau+\delta\tau}$ obtained after an infinitesimal time-step $\delta \tau$ by applying the generator $\L_{\pars({\tau})}$ to a nonequilibrium distribution $\den\tind{\tau}$ is information-theoretically closer to the equilibrium distribution than $\den\tind{\tau}$:
\begin{align}
  \Dklinf[\den\tind{\tau+\delta\tau}\| \den\rf_{\pars(\tau)}] \leq \Dklinf[\den\tind{\tau}\| \den\rf_{\pars(\tau)}].
  \label{eq:strongly-relaxing-dklinf}
\end{align}
Asymptotically, the probability distribution reaches the maximally non-committal distribution in its stationary equilibrium state.

While the strong relaxation condition may come as an appealing starting assumption for a formulation of relaxing dynamics, we still have not said anything about dynamics that implement that assumption.
In particular, we have not specified anything but time-locality for the evolution equation~\eqref{eq:evolution-operator}.
In physics, we mostly deal with evolution equations for probabilities where $\L_\pars$ is a \emph{linear operator}.
For discrete and continuous state spaces, Equation~\eqref{eq:evolution-operator} is called the Master or Fokker--Planck equation, respectively.
Such linear equations arise naturally as the equations governing the change of probability in Markov processes, in which case $\L_\pars$ is  called the (backward) generator of the Markovian dynamics.

Oliver Penrose suggested that Markov processes are ubiquitous as stochastic models for physical processes, because they ensure statistical regularity for the stochastic dynamics~\cite{Penrose1970}.
In particular, Markov processes are memoryless, which is an important criterion for the \emph{reproducibility} of statistical experiments:
All relevant information for the future evolution of the system is encoded in the instantaneous distribution.
In particular, it does not matter \emph{how} (and in which laboratory throughout the world) the state was prepared.

More precisely, for Markovian processes the conditional probability of a transition $\vec{x} \to \vec{x}'$ does only depend on the states $\vec{x}$ and $\vec{x}'$, but not on the stochastic history of the process.
Not surprisingly, this lack of memory manifests itself in information-theoretic terms:
\emph{Any} Markov process that converges to unique invariant distribution $\den^\infty$ monotonously decreases the KL--divergence of the instantaneous distribution $\den\tind{\tau}$ from $\den^\infty$:
\begin{align}
  \partial_\tau \Dklinf[\den\tind{\tau}\Vert\den^\infty] \leq 0.
  \label{eq:Markov-liapunov-general}
\end{align}
Or differently stated:
For any convergent Markov process, the relative entropy $\Dklinf[\den\tind{\tau}\Vert\den^\infty]$ is a Liapunov functional for the dynamics~\cite{Schnakenberg1976,Cover.Thomas2006}.

Keep in mind that this is a purely mathematical property.
In order to call a Markov process strongly relaxing in the present thermodynamic context, the stationary distribution $\den\rf_\pars(\vec{x})$ must originate from a physical model defined by means of extensive physical quantities $E(\vec{x};\lam)$ and $\vec{\bobs}(\vec{x})$.
More precisely, it requires that the anti-symmetric part of the generator $\L_\pars$ must be obtained as the gradient of a phase space function
\begin{align}
  \vartheta(\vec{x};\lam,\ip)=\beta(E(\vec{x},\lam) - \ip\vec{\bobs}(\vec{x}))
  \label{eq:enthalpy-observable}
\end{align}
subject to a global noise level characterized by a noise parameter $\beta$, see also \cite{Baiesi.Maes2013,Wachtel.etal2015}.
This requirement implies the condition of detailed balance, \ie the absence of probability currents in the stationary state, and thus forward and backward trajectories are equally likely.
In mathematical terms, detailed balance requires the generator $\L_\pars$ to be similar to a symmetric operator and thus to have a real spectrum.

While a consistent thermodynamics can also be formulated for Markovian dynamics that do not fulfil detailed balance, a consistent interpretation in that case requires a careful analysis of the notions of \emph{local equilibrium} and \emph{local detailed balance}~\cite{Esposito.VandenBroeck2010,Esposito2012,Seifert2011}, which are beyond the scope of this paper.

For completeness, we also mention that Liapunov functionals resembling relative entropies exist also for non-linear evolution equation, see for instance Refs.~\cite{Markowich.Villani2000,Rao.Esposito2016}.
For the purpose of the present article it is sufficient to keep in mind that any physical Markovian dynamics converging to a unique steady state constructed from a physical equilibrium potential fulfils the strong relaxation criterion:
Independently of the exact instantaneous state, they always approach the maximally uncertain equilibrium distribution that is consistent with the macroscopic thermodynamic parameters.

\subsection{Nonequilibrium Thermodynamics}
\label{sec:noneq-td}
With the thermodynamic, information-theoretic and dynamic prerequisites introduced in the previous sections, we have everything in place to proceed to formulate our nonequilibrium thermodynamics for driven and relaxing systems.
We start with the definition of nonequilibrium state variables and potentials and then consider their dynamics.
After that, we introduce process quantities like (rates of) work and heat, which allows us to define entropy production and finally formulate the Second Law of Thermodynamics.

\subsubsection{Nonequilibrium state variables and potentials, exergy}

State variables are obtained as the expectation values of phase space functions $\fct(\vec{x};\pars)$ which may explicitly depend on thermodynamic system and environment parameters $\pars$.
While we exclude phase space functions that depend explicitly on time, we explicitly include the time-dependence of state functions for driven systems where $\dot \pars = \frac{\df}{\df \tau}\pars(\tau)$ may be non-zero.
In previous sections, we have only considered equilibrium state variables (\cf~Eq.~\eqref{eq:average-ref}), that were obtained as equilibrium (reference) averages.
From now on, we will decorate such reference averages by the symbol $\star$:
\begin{align}
  \Fct_\pars\rf  := \eav{\fct}\rf :=\int\fct(\vec{x};\pars)\,\den\rf_\pars (\vec{x}) \df \vec{x}.
  \label{eq:reference-average}
\end{align}
Non-equilibrium state functions are obtained by averaging over the (explicitly time-dependent) nonequilibrium distribution $\den\tind{\tau}$:
\begin{align}
  \Fct_\pars\tind{\tau}  := \eav{\fct}\tind{\tau} :=\int\fct(\vec{x};\pars)\,\den\tind{\tau} (\vec{x}) \df \vec{x}.
  \label{eq:nonequilibrium-average}
\end{align}
While reference averages always depend on $\pars$, non-equilibrium averages $\Fct\tind{\tau} = \eav{\fct}\tind{\tau}$ depend only on $\pars$ if the phase space function $\fct(\vec{x})$ does.
For instance, the non-equilibrium averages of the macroscopic extensive observables $\obs(\vec{x})$
\begin{align}
  \Obs\tind{\tau}:=\eav{\obs}\tind{\tau}= \int \obs(\vec{x})\, \den\tind{\tau}(\vec{x})\df\vec{x}
  \label{eq:noneq-ext-obs}
\end{align}
do not depend on the control parameters.
Keep also in mind that the non-equilibrium internal energy
\begin{align}
  \U_\lam\tind{\tau} := \eav{E}\tind{\tau} = \int E(\vec{x};\lam)\, \den\tind{\tau}(\vec{x})\df\vec{x}
  \label{eq:noneq-internal-energy}
\end{align}
depends only on the \emph{system} control parameters $\lam$, but not on the intensive parameters of the environment.
In the following, we make the parameter dependence explicit, unless we find it (notationally) convenient to simply write $\Fct\rf$ instead of $\Fct\rf_\pars$.
In case we need to make the time dependence of the parameters explicit, we may also write $\Fct\rf_{\pars(\tau)}$ and $\Fct\tind{\tau}_{\pars(\tau)}$.

The non-equilibrium (information) entropy  characterizes the uncertainty of the instantaneous nonequilibrium distribution $\den\tind{\tau}$:
\begin{align}
  \Ent\tind{\tau}:=\sent[\den\tind{\tau}]=-\int\den\tind{\tau}(\vec{x})\ln\den\tind{\tau}(\vec{x}) \df \vec{x}.
  \label{eq:system-entropy}
\end{align}
In analogy to the general Legendre duality \eqref{eq:general-Legendre-duality}, we define the nonequilibrium variant of the natural dimensionless potential:
\begin{align}
  \fent_\pars\tind{\tau} :=  \beta(\U_\lam\tind{\tau} - \ip\Vobs\tind{\tau}) - \Ent\tind{\tau}
  \label{eq:noneq-potential}
\end{align}
The corresponding nonequilibrium natural energy accordingly reads
\begin{align}
  \G_\pars\tind{\tau} := \beta^{-1}\fent_\pars\tind{\tau} =  \U_\lam\tind{\tau} - \ip\Vobs\tind{\tau} - \beta^{-1}\Ent\tind{\tau}.
  \label{eq:general-noneq-potential}
\end{align}
It is straightforward to show that the difference between the natural non-equilibrium potential and its equilibrium reference yields the thermodynamic distance~\eqref{eq:dklref}:
\begin{align}
  \fent\tind{\tau}_{\pars} - \fent\rf_{\pars}=\Dkl\tind{\tau}_{\pars} \geq 0.
  \label{eq:DKL-potential}
\end{align}

By multiplying this difference with the temperature $\beta^{-1}$, one obtains a quantity with the dimensions of energy:
\begin{align}
  \B\tind{\tau}_{\pars} :=  \G\tind{\tau}_{\pars} - \G\rf_{\pars}\equiv \beta^{-1} \Dkl\tind{\tau}_{\pars} \geq 0.
  \label{eq:exergy}
\end{align}
This quantity has a strong resemblance to the quantity commonly called \emph{exergy} in an engineering (and systems theory) context~\cite{Lozano.Valero1993,Dincer.Rosen2012}.
The exergy of an equilibrium system vanishes, while it is positive for non-equilibrium distributions.
It is the maximum work that can be extracted from a system during a relaxation process.
For relaxing dynamics in the sense of Eq.~\eqref{eq:strongly-relaxing}, we directly follow that exergy is destroyed during thermodynamic relaxation:
\begin{align}
  \frac{\partial}{\partial \tau} \B\tind{\tau} \leq 0
  \label{eq:exergy-degradation}
\end{align}
The law of exergy degradation~\eqref{eq:exergy-degradation} is usually derived as a reformulation of the Second Law.
Starting from the information-theoretic perspective, it appears naturally from the start.
Its equivalence with the Second Law will become evident shortly.

\subsubsection{Time derivatives of state variables}

The reference distribution $\den\rf_\pars$ does not depend explicitly on time.
Thus the time-derivative of an equilibrium average is only due to changes in the control parameters~$\pars$:
\begin{align}
  \dot{\Fct}_{\pars}\rf := \frac{\df}{\df\tau} \Fct\rf_{\pars(\tau)} = \left(\nabla_\pars \Fct_\pars\rf\right) \dot{\pars}.
  \label{eq:dunno}
\end{align}
The parameter gradient $\nabla_\pars$ is a vector containing the partial derivatives with respect to the control parameters.
For example, in the grandcanonical situation described above we have $\nabla_\pars = (\partial_\beta,\partial_\mu,\partial_V, \partial_h)$.
For equilibrium averages, the parameter gradient
\begin{align}
  \nabla_\pars\Fct\rf = \eav{\nabla_\pars\, \fct}\rf + \cov[\fct, \nabla_\pars \vartheta]\rf.
  \label{eq:parameter-gradient}
\end{align}
can be re-written as the average of the state function $\nabla_\pars\, \fct(\vec{x})$ and an equilibrium co-variance
\begin{align}
  \cov[\fct,\fct']\rf = \eav{\fct\, \fct'}\rf - \eav{\fct}\rf\eav{\fct'}\rf.
  \label{eq:covariance}
\end{align}
The second argument in the co-variance, $\nabla_\pars \vartheta(\vec{x};\pars)$, is the gradient of the phase space function \eqref{eq:enthalpy-observable} appearing in the exponential of the reference density (\cf Eq.~\eqref{eq:general-maxEnt-distribution}).
Below, we will see its interpretation as a natural (nondimensional) state-function describing the heat content of a system.

If a state function $\fct(\vec{x})$ does not explicitly depend on the parameters, the first term in Eq.~\eqref{eq:parameter-gradient} vanishes and the co-variance is the only contribution.
In particular, we obtain the equilibrium stability relations
\begin{align}
  \frac{\partial \Obs_k\rf}{\partial \gamma_k} = \beta\cov[\obs_k,\obs_k] \geq 0.
  \label{eq:stability}
\end{align}

In contrast, the total time derivative of a non-equilibrium state variable $\Fct_\pars\tind{\tau}$ reads
\begin{align}
  \dot{\Fct}_\pars\tind{\tau} := \frac{\df}{\df \tau} \Fct\tind{\tau}_{\pars(\tau)} = \partial_\tau \Fct\tind{\tau}_\pars + (\nabla_\pars \Fct_\pars \tind{\tau}) \dot{\pars}.
  \label{eq:total-time-derivative}
\end{align}
It generally features two terms:
The first one,  which we write as a partial time-derivative, can be attributed to the (autonomous) relaxation of the non-equilibrium probability distribution:
\begin{align}
  \partial_\tau \Fct\tind{\tau}_\pars := \int \fct(\vec{x};\pars(\tau))\,\dot{\den}\tind{\tau}(\vec{x})\,\df\vec{x},
  \label{eq:general-partial}
\end{align}
For parameter-dependent observables $o(\vec{x};\pars)$ we have also the gradient part
\begin{align}
  (\nabla_\pars  \Fct_\pars \tind{\tau})\,\dot{\pars} =  \eav{\nabla_\pars\, \fct}\dot{\pars},
  \label{eq:non-equilibrium-gradient-part}
\end{align}
which cannot easily be decomposed like we did in Eq.~\eqref{eq:dunno}.

\subsubsection{Process quantities: work and heat}
Apart from changes of state variables, thermodynamics deals with process quantities that generally cannot be written as the time-derivative of a state function.
In traditional thermodynamics, process quantities are associated with inexact differentials often denoted by the symbols $\delta$ or \dj.
Here, Eq.~\eqref{eq:evolution-operator} provides us with an explicit evolution rule for the non-equilibrium probability density distribution and we do not need to work with inexact differentials.

A process quantity $\Psi\tinv{\tau}$ is defined as the integral over the time interval $[0,\tau]$ of a rate function $\dot{\Psi}\tind{\tau}$
\begin{align}
  \Psi\tinv{\tau} := \int_0^\tau \dot{\Psi}\tind{t} \df t.
  \label{eq:process-quantity-integral.}
\end{align}
In general, the integrands $\dot{\Psi}\tind{\tau}\equiv\frac{\df}{\df\tau}\Psi\tinv{\tau}$ may contain the control parameters $\pars$ as well as their derivatives $\dot{\pars}$.
In what follows, we will usually work with the integrands $\dot{\Psi}\tind{\tau}$ and not with their integrals.
However, we will see that for systems in environments of constant temperature, some process quantities for reference processes become integrable.

The central process quantities in thermodynamics are the rate of heat flow $\dot{Q}$ and the power $\dot{W}$, \ie the rate of work being performed on the system.
Work and power are attributed to changes in macroscopic extensive system control parameters $\lam$ and system observables $\Vobs$. 
Because of the distinction between control parameters and measurable observables, the rate of work per unit time (\ie, the power) $\dot{W}\tind{\tau}$ consists of two terms:
\begin{align}
  \dot{W}\tind{\tau} := \dot{W}\tind{\tau}\drv + \dot{W}\tind{\tau}\conv ,
  \label{eq:work-rate}
\end{align}
which we will call the \textit{driving} and \textit{autonomous} power.
Both work terms have a positive sign if they add a positive amount of energy to the system.

The driving power is the rate of energy change obtained by changing the control parameters that appear in the energy function of the system:
\begin{align}
  \dot{W}\tind{\tau}\drv &:= \eav{\frac{\partial E_\lam}{\partial\lam}}\tind{\tau}\dot{\lam}\,= \sum_i \R_i\tind{\tau} \dot{\lambda}_i\label{eq:driving-work} \quad\\
  \big[\,&=  - p\tind{\tau}\dot{V} + \pol\tind{\tau}\dot{h}\,\big]\nonumber,
\end{align}
The exact form of the driving power depends on the physical situation.
For instance, in the grandcanonical case where volume and (and potentially a field $h$) is controlled, the driving work contains the mechanical work (and the change in the internal field energy due to a change in the bare field $h$).
Like in equilibrium, the quantities 
\begin{align}
  \R_i\tind{\tau} = \eav{\frac{\partial E}{\partial \lambda_i}}\tind{\tau}
  \label{eq:non-equilibrium-internal-obs}
\end{align}
 are the non-equilibrium system observables that are conjugate to the system control parameters $\lambda_i$ and can be either intensive or extensive, \cf Eq.~\eqref{eq:system-conjugate-qantities}.

The autonomous power is the rate of energy that enters (positive sign) or leaves (negative sign) the system as a result of the autonomous system dynamics~\eqref{eq:evolution-operator}.
It amounts to the exchange of conserved extensive quantities $\Vobs\tind{\tau}$ with the reservoirs against their intensive parameters $\ip$:
\begin{align}
  \dot{W}\tind{\tau}\conv := \ip \vec{\dot{\Obs}}\tind{\tau}\quad \big[\,= \mu \dot{\N}\tind{\tau}\big].
  \label{eq:autonomous-work}
\end{align}
It is the power that can be extracted from the non-equilibrium state of the system by \emph{consuming exergy} that was, for instance, generated from driving the system out of an equilibrium state.
Like before, the equality in square brackets holds for the grandcanonical ensemble where the autonomous work is due to the autonomous flow of matter between the system and the environment.

In standard thermodynamics, work contributions are usually split into \emph{input} and \emph{output} work, and the distinction is made by the sign.
Our distinction relies on whether extensive quantities are either controlled or may autonomously relax against an environment.
Macroscopic well-known thermodynamic process (\eg~the Carnot-, Otto or Diesel-processes) are constructed from combining four different processes where different system and environment parameters are controlled.
Often, the work during one of this partial processes has a well-defined sign, and is thus either a pure input or a pure output work.
Using the present distinction we still may find a unique attribution of the work along a partial process to be either of the driving or the autonomous type.
Notice however, that reversing the thermodynamic cycle reverses the sign of the power.
It thus exchanges input with output work, whereas the distinction into driving or autonomous work remains the same.

\subsubsection{The First and Second Law of Thermodynamics}

Besides through work, the internal energy of a system may change due to a heat flow $\dot{Q}\tind{\tau}$ from the environment to the system.
The First Law of Thermodynamics expresses energy conservation and states that the internal energy of system can change either by a heat flow or through power:
\begin{align}
  \dot{\U}\tind{\tau} = \dot{Q}\tind{\tau} + \dot{W}\tind{\tau}.
  \label{eq:first-law}
\end{align}
Notice that the First Law and the definition of work unambiguously define the rate of heat flow:
It is the rate of change of the energy of the system, that cannot be attributed to changes in macroscopic properties that we use to specify equilibrium state of the system.

If the environment is at an inverse temperature $\beta$, the heat flow corresponds to an entropy flow $\dot{\Ent}\tind{\tau}_\text{e}$ from the environment to the system:
\begin{align}
  \dot{\Ent}\tind{\tau}_\text{e} &:= \beta\dot{Q}\tind{\tau} = \beta(\dot{\U}\tind{\tau} - \dot{W}\tind{\tau}) \label{eq:entropy-flow} \\
  \nonumber\big[\, &=\beta( \dot{\U}\tind{\tau} -\mu\dot{\N}\tind{\tau} + p\tind{\tau}\dot{V} - \pol\tind{\tau}\dot{h}  )\,\big].
\end{align}
The interaction of the system with its environment changes the nonequilibrium entropy, \ie the instantaneous uncertainty of the distribution.
However, it also changes the true microscopic state of the environment by means of introducing correlations between the system and the environment.
The latter are not part of our description and lead to an irreversible production of entropy~$\dot{\Ent}\tind{\tau}_\text{i}$.

In the present work, this entropy production emerges as the change of system entropy that cannot be attributed to an entropy flow $\dot{\Ent}\tind{\tau}_\text{e}$:
\begin{align}
  \dot{\Ent}\tind{\tau}_\text{i} := \dot{\Ent}\tind{\tau} - \dot{\Ent}\tind{\tau}_\text{e}. \label{eq:entropy-production}
\end{align}
Keep in mind that although the time-integrals $S_\text{i}\tinv{\tau}$ and $S_\text{e}\tinv{\tau}$ are well defined, there is no corresponding state observable for these entropy changes.
Using the definitions of heat~\eqref{eq:first-law} and autonomous power~\eqref{eq:autonomous-work}, we can rewrite the irreversible entropy production as
\begin{align*}
  \dot{\Ent}\tind{\tau}_\text{i} &=  \dot{\Ent}\tind{\tau} -\beta(\dot{\U}_\lam\tind{\tau} - \dot{W}\tind{\tau})\\
  &=  \dot{\Ent}\tind{\tau} -\beta\dot{\U}_\lam\tind{\tau} + \beta \dot{W}\tind{\tau}\conv + \beta\dot{W}\tind{\tau}\drv \\
  &=  \dot{\Ent}\tind{\tau} -\beta(\dot{\U}_\lam\tind{\tau} - \ip \vec{\dot{\Obs}}\tind{\tau}) + \beta\dot{W}\tind{\tau}\drv.\\
\end{align*}

In order to formulate the usual notion of the Second Law, we consider the gradient part of time-derivative for the natural non-equilibrium potential~\eqref{eq:noneq-potential}, which we call the gradient entropy flow 
\begin{align}
  \dot{\Ent}_\nabla\tind{\tau} &:=  (\nabla_\pars \fent_\pars\tind{\tau})\dot{\pars}\\
  &= (\U_\lam\tind{\tau} - \ip\Vobs\tind{\tau})\dot{\beta} - \beta\Vobs\tind{\tau}\dot{\ip} + \beta\frac{\partial{\U_\lam\tind{\tau}}}{\partial\lam}\dot{\lam}\nonumber\\
  &= (\U_\lam\tind{\tau} - \ip\Vobs\tind{\tau})\dot{\beta} - \beta\Vobs\tind{\tau}\dot{\ip} + \beta\dot{W}\tind{\tau}\drv.
  \label{eq:gradient-part-noneq}
\end{align}
Using partial integration, we realize that the irreversible entropy production 
\begin{align}
 \dot{\Ent}\tind{\tau}_\text{i} 
  &=  -\dot{\fent}_\pars\tind{\tau} +({\U}_\lam\tind{\tau} - \ip \vec{{\Obs}}\tind{\tau})\dot{\beta} -\beta\Vobs\tind{\tau}\dot{\ip} + \beta\dot{W}\tind{\tau}\drv \nonumber\\
  &\equiv -\dot{\fent}_\pars\tind{\tau} + \dot{\Ent}\tind{\tau}_\nabla\nonumber\\&
  = -\dot{\fent}_\pars\tind{\tau} + (\nabla_\pars \fent_\pars\tind{\tau})\dot{\pars}
\end{align}
is nothing than the autonomous relaxational part of the evolution of the natural potential, represented by the partial time-derivative
\begin{align}
  \dot{\Ent}\tind{\tau}_\text{i}  &= -\partial_\tau {\fent_\pars}\tind{\tau}.
  \label{eq:irrev-ep-as-relaxing-noneq-potential}
\end{align}

In order to formulate the common statement of the Second Law of Thermodynamics, we use that the natural reference potential $\fent_\pars\rf$ does not explicitly depend on time, \ie that $\partial_\tau\fent_\pars\rf=0$.
Hence, we can equally express the irreversible entropy production using the thermodynamic distance~\eqref{eq:DKL-potential}:
\begin{align}
  \dot{\Ent}\tind{\tau}_\text{i}   &= -\partial_\tau [\fent\tind{\tau}_\pars - \fent\rf_\pars]= -\partial_\tau \Dkl_\pars\tind{\tau}.
  \label{eq:irrev-ep-as-relaxing-DKL}
\end{align}

Equation~\eqref{eq:irrev-ep-as-relaxing-DKL} is the main result of this paper.
It establishes a fundamental relation between the definition of thermodynamic entropy production (appearing on the left hand side) and the abstract definition of an information-theoretic thermodynamic distance (on the right hand side).
While structurally similar results have appeared in various more specialized contexts~\cite{Procaccia.Levine1976,Landauer1961,Jarzynski1997,Vaikuntanathan.Jarzynski2009,Takara.etal2010,Esposito.VandenBroeck2011}, the present general derivation is the main original conceptional result of this work.
In particular, we emphasize that we didn't rely on any implicit assumptions.
The thermodynamic entropy production is defined \emph{operationally} and consistent with phenomenological thermodynamics:
it originates from the (physically unambiguous) definition of power, Eq.~\eqref{eq:work-rate}, which specifies energy changes due to changes in experimentally accessible, \ie measurable and controllable, macroscopic quantities.
In contrast, the right-hand side of~\eqref{eq:irrev-ep-as-relaxing-DKL} has an unambiguous information-theoretic meaning:
it measures the instantaneous rate of discarded information with respect to a physical equilibrium reference measure.

The above definitions do not rely on the strong relaxation condition~\eqref{eq:strongly-relaxing}.
Yet, assuming strongly relaxing dynamics, Eq.~\eqref{eq:irrev-ep-as-relaxing-DKL} directly yields the Second Law in its most common form:
\begin{align}
   \dot{\Ent}\tind{\tau}_\text{i} &=- \partial_\tau \Dkl_{\pars}\tind{\tau} \geq 0.
   \label{eq:second-law}
\end{align}
The framework presented in this work does provides a \emph{proof} of the Second Law using only strong relaxation as an assumption.
In particular, we proved the Second Law for arbitrary Markovian dynamics that, in the absence of driving, relax to a thermodynamic equilibrium distribution.
To the knowledge of the author, this is the most general derivation of this fundamental law based on first principles.

\subsection{Derived concepts}
\label{sec:derived-concepts}
So far we have illustrated the connection between information theory and the elementary notions of thermodynamics in a very general way.
The present section focuses on generalizations of other derived thermodynamic concepts like enthalpy.
We further derive and discuss other (equivalent) thermodynamic inequalities including Landauer's principle of information-to-work conversion.

\subsubsection{Enthalpy as heat content and mediate work}
Another important concept in traditional thermodynamics is that of enthalpy which arises naturally in the context of chemical physics.
In traditional (bulk) chemistry it is natural to work with the isobaric-isothermal ensemble, where pressure and temperature of the environment are controlled and volume can fluctuate.
The equilibrium enthalpy $\H\rf = \U\rf+p\V\rf$ can be interpreted as the energy that is necessary to create a system minus the amount of work performed \emph{by} the system (notice the sign) to establish its volume in a given environment.
Thus, enthalpy can be understood as the ``heat content'' of a system, \ie the part of the internal energy that is not attributed to energy in macroscopic degrees of freedom.

It seems natural to generalize the notion of heat content to our ensembles for general open systems.
For the case of a system with autonomous (fluctuating) particle numbers (and all other extensive parameters constant), the analogous equilibrium heat content reads $\U\rf -\mu\N\rf$:
It is the energy needed to create a system plus the amount of work needed to establish a certain particle number within a large environment characterized by a chemical potential $\mu$.

For the general case, we thus define the natural ``non-free'' potential
\begin{align}
  \Theta := \beta\U -\beta\ip\Vobs = \eav{\vartheta}.
  \label{eq:theta-potential}
\end{align}
The natural non-free potential is the dimensionless version of what we call the natural heat content or generalized enthalpy
\begin{align}
  \H := \beta^{-1}\Theta = \U -\ip\Vobs.
  \label{eq:generalized-enthalpy}
\end{align}
The heat content thus emerges as an energetic ``non-free'' state function:
It specifies an energy content that is not available to perform work.
Notice that the natural non-free  potential appear both the Legendre transform of $\beta \U$ with respect to the observed, extensive quantities $\beta\ip\Vobs$ and the (non-equilibrium or reference) average of the state function $\vartheta$ defined in Eq.~\eqref{eq:enthalpy-observable}.

To see more clearly how  heat content and heat flow is related, consider the following situation:
A system in a given state (\ie with a given distribution $\den\tind{\tau}$) resides in an environment at temperature $\beta$ and other intensive parameters $\ip$.
The state uniquely determines the expectation values of the extensive observables $\Vobs\tind{\tau}$.
The heat content $\H\tind{\tau} = \U\tind{\tau} - \ip \Vobs\tind{\tau}$ characterizes the part of the internal energy, that we cannot attribute to macroscopic degrees of freedom $\ip\Vobs\tind{\tau}$.
Now suppose we instantaneously change the intensive parameters from $\ip$ to $\ip'$.
The internal energy of the system has not changed, but the attribution of what constitutes the macroscopic part of the energy is now given by $\ip' \Vobs\tind{\tau}$.
As such, the heat content of the system will have changed, without either heat flowing or work being performed on the system.
Instead, in order to change the parameters of the reservoirs, work will have to be performed on the environment.
Since the environment is imagined as infinitely large, it is not sensible asking for the total amount of work needed to change its state.
However, the (negative) power needed due to the instantaneous state of system is well defined:
\begin{align}
  \dot{W}\med\tind{\tau} := -\Vobs\tind{\tau}\dot{\ip}.
  \label{eq:mediate-work}
\end{align}
We suggest to call $\dot{W}\med\tind{\tau}$ the \emph{mediate power}.
We apply the negative sign convention, because energy changes are formulated with respect to the system.
The word ``mediate'' here is understood in contrast to the immediate driving power $\dot{W}\drv$, which directly changes the internal energy of the system by changing the parameters that govern the energy function of the system.
In contrast, the mediate power reflects how our distinction between energy in hidden (microscopic) and observable (macroscopic) degrees of freedom changes with time if we vary the environment parameters.
The sum of autonomous and mediate power is the change of the energy attributed to macroscopic, observable degrees of freedom:
\begin{align}
  \dot{W}\conv - \dot{W}\med = \frac{\df}{\df \tau} [\Vobs\tind{\tau}\ip] \equiv \frac{\df}{\df \tau}[{\U\tind{\tau} -\H\tind{\tau}}]
  \label{eq:total-change-in-obs-dof}
\end{align}

The mediate power appears naturally in the $\ip$-derivatives of the thermodynamic potentials.
In fact, subtracting the entropy flow $\Ent\tind{\tau}_\text{e} = -\beta \dot{Q}\tind{\tau}$ from the total change of the ``non-free'' natural potential $\dot{\Theta}$, we recover the gradient entropy flow that has appeared in~Eq.~\eqref{eq:gradient-part-noneq}:
\begin{align}
  \dot{\Theta} - \beta \dot{Q}\tind{\tau} 
  &= (\U-\ip\Vobs)\dot{\beta} - \beta\Vobs\dot{\ip} + \beta \dot{W}\drv\nonumber \\
  &= (\nabla_\pars \fent_\pars\tind{\tau})\dot{\pars}\nonumber\\
  &\equiv \dot{\Ent}_\nabla\tind{\tau}.
  \label{eq:gradient-derivative-dimless-enthalpy}
\end{align}
Notice that the corresponding dimensional relation
\begin{align}
  \dot{\H}\tind{\tau} - \dot{Q}\tind{\tau} &= (\nabla_\pars \G\tind{\tau}_\pars)\dot\pars - \beta^{-1}\H\tind{\tau}\dot{\beta}
  \label{eq:dimensional}
\end{align}
features and additional non-gradient term for non-isothermal processes.
Still, the following relation is valid for arbitrary processes:
\begin{align}
  \dot\H\tind{\tau} = \dot{Q}\tind{\tau} + \dot{W}\drv\tind{\tau} + \dot{W}\med\tind{\tau}.
  \label{eq:enthalpy-change-splitting}
\end{align}
In the above equation, we see that a change of enthalpy can be due to three physically distinct mechanisms:
Heat flow $\dot{Q}\tind{\tau}$, (immediate) driving power $ \dot{W}\drv\tind{\tau}$ and the mediate power associated to changing the intensive parameters characterizing the environment.
As such, Eq.~\eqref{eq:enthalpy-change-splitting} acts as a ``First Law'' for the (generalized) enthalpy.
Different formulations (although similar in spirit) have been proposed for chemical reaction networks in Refs.~\cite{Rao.Esposito2016,Schmiedl.Seifert2007}.

\subsubsection{Work inequalities}
For arbitrary processes between equilibrium states, a common result in traditional thermodynamics is that the work performed on the system is necessarily larger than the difference of the corresponding equilibrium free energies.
Here, we can generalize this inequality to a non-equilibrium version for driven systems in a straightforward way.
To that end notice that by definition we have
\begin{align}
  \dot{W}\tind{\tau} &\equiv \dot{\U}\tind{\tau} - \beta^{-1}\dot{\Ent}_\text{e}\tind{\tau}\nonumber\\
  &\equiv\dot{\U} - \beta^{-1}\dot{\Ent}\tind{\tau} + \beta^{-1}\dot{\Ent}_\text{i}\tind{\tau}\nonumber\\
  &\geq \dot{\U}\tind{\tau} - \beta^{-1}\dot{\Ent}\tind{\tau},
  \label{eq:work-inequality}
\end{align}
where in the last line we used the Second Law, Eq.~\eqref{eq:second-law}.
Again this inequality is \emph{equivalent} to the Second Law and thus to our initial information-theoretic assumption of relaxing dynamics.
Using the definition of the non-equilibrium free energy $\Free\tind{\tau}\equiv \U\tind{\tau} - \beta^{-1}\Ent\tind{\tau}$ we use the chain rule to find the relation
\begin{align}
  \dot{W}\tind{\tau} \geq &\,\dot{\Free}\tind{\tau} - \beta^{2}\Ent\tind{\tau}\dot{\beta}\label{eq:work-ineqality-free-energy}
  \equiv  \dot{\Free}\tind{\tau} + \Ent\tind{\tau}\dot{T} 
\end{align}
This is the non-equilibrium version of the work inequality.
In particular, integrating this relation along an isothermal process with $\dot{\beta}=0$ we have
\begin{align}
  W\tinv{\tau}\geq \Free\tind{\tau}_{\pars(\tau)} - \Free\tind{0}_{\pars(0)}.
  \label{eq:Jarzynski-inequality}
\end{align}
For any isothermal process, the work done on the system is at least as large as the difference between the non-equilibrium free energies at the beginning and at the end of the process.
For a non-isothermal process, the lower bound for the power contains a non-integrable rate term weighing the rate of temperature change $\dot{T}$ with the non-equilibrium entropy $\Ent\tind{\tau}$.

Sometimes, the work inequality~\eqref{eq:Jarzynski-inequality} is stated with the differences of the \emph{equilibrium} free energies appearing on the right-hand side.
However, it is important that such an inequality is not valid for arbitrary non-equilibrium situations.
While we have already seen that it usually requires an isothermal processes, we also need to be in a physical situation where here are no autonomous measurable macroscopic extensive observables.
Hence, the work is purely driving work and does not have an autonomous contribution.
Assuming that driving only occurs during the interval $[0,\tau]$, we have $\pars(t) = \pars_\text{f}$ and thus no driving power for $t\geq\tau$.
It follows that
\begin{align}
  W\tinv{\tau} &= W\drv\tinv{\tau} = W\drv[0,\infty]\nonumber\\
  &\geq \Free\tind{\infty}_{\pars(\infty)}- \Free\tind{0}_{\pars(0)}\nonumber\\
  &= \Free\rf_{\pars_\text{f}} - \Free\tind{0}_{\pars_\text{i}}.
  \label{eq:work-relation-equilibrium-repaired}
\end{align}
Consequently, even in the canonical isothermal case, a general inequality that involves only the equilibrium free energies requires system to \emph{start} in equilibrium~\cite{Jarzynski1997}.

This example reminds us that when using well-known thermodynamic inequalities involving difference of state function at different times, one needs to be careful.
While Eq.~\eqref{eq:work-ineqality-free-energy} is always valid, the more subtle issues arising for open systems in this formulation are implicit in the definition of work.
In the following make this more explicit by introducing the notion of reversible work.

\subsubsection{Reversible processes and reversible work}
Traditional thermodynamics uses the notion of idealized \emph{reversible} process in order to formulate thermodynamic inequalities.
The reversible work is defined as the net work that can be extracted from a system in such a reversible process, where the system remains at equilibrium at every point in time.
Note that this condition requires the process to be quasistatic, and thus would require an infinite amount of time to perform.

In order to see how reversible processes enter the present picture, we start with the definition of the reversible power.
In analogy  definitions~\eqref{eq:driving-work} and~\eqref{eq:autonomous-work}, we separate it into two terms:
\begin{align}
  \dot{W}\tind{\tau}\rev := \dot{W}\tind{\tau}\drvrv + \dot{W}\tind{\tau}\convrv,
  \label{eq:reversible-work}
\end{align}
with the reference driving power $\dot{W}\tind{\tau}\drvrv$ and the reference autonomous power $\dot{W}\tind{\tau}\convrv$ defined as
\begin{subequations}
\begin{align}
  \dot{W}\tind{\tau}\drvrv &:= \eav{\frac{\partial E_\lam}{\partial\lam}}\rf\dot{\lam}= \sum_i \R_i\rf \dot{\lambda}_i, \label{eq:rev-driving-work} \\
  \dot{W}\tind{\tau}\convrv &:= \ip \dot{\Vobs}\rf.  \label{eq:rev-autonomous-work}
\end{align}
\label{eq:rev-work-defs}
\end{subequations}
Formally, we have exchanged non-equilibrium averages by their corresponding equilibrium reference values.
Keep in mind that while $\vec{\R}\tind{\tau} = \partial_\lam\U\tind{\tau}$, the Maxwell relations~\eqref{eq:Maxwell-relations} show is that in general $\R_i\rf \neq \partial_\lam \U\rf$.
Moreover, the term ``autonomous'' power might not be appropriate.
In a reversible process, there is no autonomous relaxation by definition.
For practical purposes, the concrete expressions for $\R_i\rf$ and 
\begin{align}
  \dot{\Obs}\rf = \frac{\df}{\df \tau}\left[\Obs\rf(\pars(\tau))\right]
  \label{eq:reference-time-average}
\end{align}
are fully determined by the equilibrium state functions $\Vobs(\pars)$ describing a (traditional) thermodynamic system.
The irreversible power is defined as the difference between the actual work and its reversible analogue:
\begin{align}
  \dot{W}\tind{\tau}_\text{irr} := \dot{W}\tind{\tau}-\dot{W}\tind{\tau}\rev.
  \label{eq:irreversible-work}
\end{align}

As equilibrium state variables serve as reference values for their non-equilibrium analogue, reversible processes can be understood as (idealized) reference processes for real processes.
It is a well know fact that in an isothermal process (\ie a process, where the bath temperature remains constant ($\dot \beta = 0$) in the interval $[0,\tau]$), the difference between equilibrium free energies is given by the integrated reversible work.
In order to proof this statement for the present situation, we use Eq.~\eqref{eq:parameter-derivatives-generating} and the definitions~\eqref{eq:rev-work-defs} to find: 
\begin{align}
  \dot{W}\tind{\tau}\rev &= (\partial_{\lam}\G_\pars\rf)\dot\lam + (\partial_{\ip}\G_\pars\rf)\dot\ip   + \frac{\df}{\df \tau}\left[ \ip\Vobs\rf \right]\nonumber\\
  &=(\nabla_\pars\G_\pars\rf)\dot{\pars} - (\partial_\beta \G_\pars\rf)\dot{\beta} + \frac{\df}{\df \tau}\left[ \ip\Vobs\rf \right]\label{eq:rev-work-first-reformulation}.
\end{align}
The $\beta$-derivative of the natural potential is easily recognized as  
\begin{align}
  \partial_\beta \G_\pars\rf &= \partial_\beta( \beta^{-1} \fent_\pars\rf) = \beta^{-1}\partial_\beta\fent - \beta^{-2}\fent\rf\nonumber\\
  & =\beta^{-1}(\U\rf-\ip\Vobs\rf) - \beta^{-2}(\beta \U\rf -\beta\ip\Vobs\rf-\Ent\rf)\nonumber\\
  &= \beta^{-2}\Ent\rf .
  \label{eq:beta-derivative-dimensional-potential}
\end{align}
Realizing the first term of Eq.~\eqref{eq:rev-work-first-reformulation} as the total time-derivative of the natural equilibrium energy $\G\rf_\pars$, we thus find
\begin{align}
  \dot{W}\tind{\tau}\rev
  &= \frac{\df}{\df \tau}\left[\G + \ip\Vobs\rf \right] -(\beta^{-2}\Ent\rf)\dot{\beta} \nonumber\\
  &= \dot{\Free}\rf + \Ent\rf\dot{T},
  \label{eq:rev-work-rewrite}
\end{align}
We immediately recognize the structural similarity of the \emph{equality}~\eqref{eq:rev-work-rewrite} for the reversible work with the \emph{inequality}~\eqref{eq:work-ineqality-free-energy} for the actual work.
By integration, we obtain the traditional result mentioned above for isothermal processes:
\begin{align}
  W\rev\tinv{\tau}
  &= \Free\rf_{\pars(\tau)} - \Free\rf_{\pars(0)}.
  \label{eq:free-energy-difference}
\end{align}
For non-isothermal processes we cannot integrate Eq.~\eqref{eq:rev-work-rewrite} and thus have no direct relation between reversible work and free energy changes.

\subsubsection{Thermodynamics and information processing}
While in the previous section we derived some connections between work and free energy, our main result~\eqref{eq:irrev-ep-as-relaxing-DKL} allows us to relate work and information.
Originally, this question was raised by Landauer regarding the thermodynamic cost of irreversible computation~\cite{Landauer1961}.

Landauer's idea is summarized as follows:
Consider the state of a single \emph{bit}, \ie a single binary digit of information, as part of the microscopic state of the system.
Suppose that there is no energy bias associated with the state of this bit.
Without any a-priori knowledge about the bit, it will be in one of its two states (say \zero or \one) with equal probability.
We thus have $p_{\zero} = p_{\one}=0.5$ with an associated information entropy $S_\text{bit} = \ln{2}=1\mathrm{bit}$.
The elementary irreversible single-bit operation is deletion, \ie the reset of the bit to a reference, say $\one$.
After erasure, the bit is certainly ($p_{\one}=1$) in state $\one$ and the entropy of the bit vanishes.
Thus, the erasure of a bit leads to a \emph{decrease} in system entropy by the amount $-\ln{2}$.
By the Second Law, the irreversibly produced entropy should be positive.

Landauer's principle is the statement that the erasure of a bit at temperature $\beta^{-1}$ comes at thermodynamic cost of at least $Q_\text{L} := \beta^{-1}\ln{2}$ of dissipation.
The energy for this dissipation must enter the system as an amount of work $W^{\to\one}$.
Consequently, the work necessary for the erasure of a single bit at constant temperature obeys the inequality
\begin{align}
  W^{\to\one} \geq Q_\text{L} = \beta^{-1}\ln{2}.
  \label{eq:Landauer-principle-erasure}
\end{align}

We can easily derive Landauer's principle from the isothermal work inequality~\eqref{eq:Jarzynski-inequality}.
Consider an arbitrary isothermal process in the time interval $[0,\tau]$, where at the initial time $t=0$ the bit is in state $\one$ and $\zero$ with equal probability, and at the final time $t=\tau$ is is surely in state $\one$.
If this is the only effect of the process, the difference in nonequilibrium entropies is $\Ent\tind{\tau} - \Ent\tind{0} = -\ln 2$.
Now assume that the bit is energetically neutral, meaning the internal energy does not depend on its state ($\U\tind{\tau} = \U\tind{0}$) and thus
\begin{align}
  W\tinv{\tau} &\geq \Free\tind{\tau} -\Free\tind{0}\nonumber\\ 
  &= \U\tind{\tau} - \U\tind{0} - \beta^{-1}(\Ent\tind{\tau} - \Ent\tind{0})\nonumber\\
    &=\beta^{-1}\ln2,
  \label{eq:Landauer-erasure-derived}
\end{align}
which is exactly the statement of the original formulation~\eqref{eq:Landauer-principle-erasure}.

Recently, related inequalities have been obtained in various contexts~\cite{Vaikuntanathan.Jarzynski2009,Takara.etal2010,Esposito.VandenBroeck2011,Rao.Esposito2016}, which relate the irreversible work to the change in the thermodynamic distance Eq.~\eqref{eq:dklref}.
To connect to these formulations, it will be convenient to define the shorthand notation
\begin{align}
  \dref{\tau}\Fct := \Fct\tind{\tau}_{\pars(\tau)} - \Fct\rf_{\pars(\tau)}
  \label{eq:difference-operator}
\end{align}
for the differences of non-equilibrium state function $\Fct\tind{\tau}$ and their corresponding equilibrium reference value $\Fct\rf$.
Subtracting Eq.~\eqref{eq:rev-work-rewrite} from the work inequality \eqref{eq:work-ineqality-free-energy} and multiply by $\beta$ to find
\begin{align}
  \beta \dot{W}\tind{\tau}_\text{irr} \geq\, & \beta \dref{\tau}\dot{\Free} - \beta^{-1}(\dref{\tau}\Ent)\dot{\beta}\nonumber\\
  =\,&\beta \dref{\tau}\dot{\G} + \beta \frac{\df}{\df \tau}[\ip\dref{\tau}\Vobs] -\beta^{-1}(\dref{\tau}\Ent)\dot{\beta}, \label{eq:derivation-Landauer-differences}
\end{align}
where we used the symbol $\dref{\tau}$ defined in Eq.~\eqref{eq:difference-operator} for brevity of notation.
The chain rule lets us express the first term using the total time-derivative of the thermodynamic distance~\eqref{eq:DKL-potential}:
\begin{align}
  \beta \dref{\tau}\dot{\G} &= \dref{\tau}\dot{\fent} - (\dref{\tau}\G)\dot{\beta}\nonumber\\
  &= \dot{\Dkl}\tind{\tau}_\pars - (\dref{\tau}\G)\dot{\beta}.
  \label{eq:derivation-Landauer-DKL}
\end{align}
Inserting Eq.~\eqref{eq:derivation-Landauer-DKL} into Eq.~\eqref{eq:derivation-Landauer-differences} while realizing that $\dref{\tau}[\G + \beta^{-1}\Ent] = \dref{\tau}{\H}$ we find the general inequality
\begin{align}
  \beta \dot{W}\tind{\tau}_\text{irr} \geq \dot{\Dkl}\tind{\tau}_\pars +  \beta \frac{\df}{\df \tau}[\ip\dref{\tau}\Vobs]- (\dref{\tau}\H)\dot{\beta}.
  \label{eq:landauer-with-dkl}
\end{align}
While this result is equivalent to Eq.~\eqref{eq:irrev-ep-as-relaxing-DKL}, we consider it as another main result of this work.
It is valid for arbitrary driven processes, as long as the dynamics is strongly relaxing.
In order to get a better intuition for the terms appearing on the right-hand side of inequality~\eqref{eq:landauer-with-dkl}, we consider some special cases.

\subsubsection{Process-specific bounds for the irreversible work}
First consider \emph{isothermal} processes where also all extensive quantities are controlled.
Then, we have no autonomous extensive observables $\vec{\bobs}(\vec{x})$ or conjugate intensive parameters $\ip$.
The only remaining term in the bound is the thermodynamic distance and the bound reads
\begin{align}
   \beta \dot{W}\tind{\tau}_\text{irr} \geq \dot{\Dkl}\tind{\tau}_\pars.
  \label{eq:noneq-Landauer-principle-esposito}
\end{align}
The right hand side can be integrated to reproduce the ``nonequilibrium Landauer principle'' from Ref.~\cite{Esposito.VandenBroeck2011}:
\begin{align}
  W_\text{irr}\tinv{\tau} \geq \Dkl\tind{\tau}_{\pars(\tau)} - \Dkl\tind{0}_{\pars(0)}
  \label{eq:noneq-Landauer-integrated-esposito}
\end{align}
It states that in a canonical isothermal scenario the irreversible entropy production is at least as large as the difference of the thermodynamic distances at the end and the beginning of the process.

For general open isothermal systems the bound reads
\begin{align}
   \beta \dot{W}_\text{irr}\tind{\tau} \geq \dot{\Dkl}\tind{\tau}_\pars +  \beta \frac{\df}{\df \tau}[\ip\dref{\tau}\Vobs].
  \label{eq:noneq-Landauer-principle-isothermal-general}
\end{align}
The additional term contains the rate of change of the deviation of instantaneous energy attributed to macroscopic degrees of freedom, $\ip\Vobs\tind{\tau}$, from its value in the corresponding equilibrium state, $\ip\Vobs\rf$.
It can still be integrated to yield
\begin{align}
  W_\text{irr}\tinv{\tau} \geq \Dkl\tind{\tau}_{\pars(\tau)} - \Dkl\tind{0}_{\pars(0)}
  +\ip(\tau)\dref{\tau}\Vobs - \ip(0)\dref{0}\Vobs.
  \nonumber
\end{align}
For an isothermal process where the observable macroscopic extensive quantities are at their equilibrium values at the beginning and the end of the process, we reproduce the bound from Eq.~\eqref{eq:noneq-Landauer-principle-isothermal-general}.
Notice that this result does not depend on the value of this difference at any point \emph{during} the process.
Moreover, it does not require the system to be in equilibrium at the beginning or the end of the process.
We only required that the \emph{observable} extensive quantities are at their equilibrium values, irrespective of the value of the internal energy or the exact form of the distributions.

For non-isothermal process, the general lower bound for the irreversible work, Eq.~\eqref{eq:landauer-with-dkl}, cannot be integrated.
The bound thus depends on the value of all observables and parameters during the process.
In particular, unlike for the integrable case, it will crucially depend on the speed of the driving and its total duration.
Disregarding the question whether such processes exist, we can still formulate the notions of quasi-static processes.

A \emph{phenomenologically quasi-static process} is the limit of a process such that at any instant during the duration of the process we have $\dref{t}\Vobs = 0$.
The bound for such a process is formally identical to the bound in a process for closed systems, \ie in physical situation where there are no autonomous macroscopic extensive observables.
In both situations, only the internal energy can deviate from its equilibrium value and thus $\dref{t}\H = \dref{t}\U$.
The inequality for closed systems and phenomenologically quasi-static processes thus reads
\begin{align}
   \beta \dot{W}\tind{\tau}_\text{irr} \geq \dot{\Dkl}\tind{\tau}_\pars - (\dref{\tau}\U)\dot{\beta}.
  \label{eq:noneq-Landauer-principle-canonical-isothermal}
\end{align}
We further define a \emph{full quasi-static process} as a limiting process where in addition to $\Vobs$ also the internal energy $\U$ is always at equilibrium.
In that case, we recover the isothermal canonical bound~Eq.~\eqref{eq:noneq-Landauer-principle-esposito}.
Notice that while a reversible process by definition is a full quasi-static process, the reverse is not true.
A (full) quasistatic process constrains only the average values of the extensive quantities during the process, whereas a reversible process specifies the entire instantaneous distribution as $\den\tind{t} = \den\rf_{\pars(t)}$.

\subsection{Discussion}
\label{sec:discussion}
In the previous two sections, we showed how thermodynamics inequalities can be derived in a self-consistent information-theoretic framework based on basic physical concepts.
For a reader that has been following the recent developments in non-equilibrium thermodynamics, the validity of these results may not come as a huge surprise.
In particular, some notions of work and heat in open (but non-driven) systems has been analysed in Ref.~\cite{Procaccia.Levine1976}.

While the present work is more general and contains a full treatment of driven systems, its main purpose is to clarify the logical structure of thermodynamics.
In particular, we tried to be very clear on disentangling the different elementary building blocks entering this presentation:
\begin{itemize}
  \item The basic concept of physics and classical mechanics is energy.
    Equilibrium statistical mechanics relates macroscopic averages of a microscopic energy function $E(\vec{x})$ and other (observable) extensive quantities $\vec{\bobs}(\vec{x})$.
    The theory of large deviations ensures that these extensive averages capture the typical behaviour (Sections \ref{sec:canonical-ensemble} and \ref{sec:generalized-ensembles}).
  \item Probability theory formalizes plausible reasoning; information theory allows us to quantify and compare the uncertainty of thermodynamic states formalized as statistical ensembles.
    Applying these notions to physical systems, we obtain the MaxEnt principle (Sec.~\ref{sec:information-theory}).
  \item Markov processes discard information and thus formalize abstract irreversibility.
    Strongly relaxing dynamics connect this notion of irreversibility to the physical irreversibility of the macroscopic relaxation process (Sec.~\ref{sec:relaxing-dynamics}).
  \item Apart from energy, which is the fundamental physical quantity, all other relevant phase-space functions are either physical parameters or observable quantities.
    Consequently, defining non-equilibrium state functions does not extend the physical description (Sec.~\ref{sec:noneq-td}).
  \item The connection to traditional thermodynamics only requires the First Law, Eq.~\eqref{eq:first-law}, which expresses energy conservation between macroscopic and microscopic forms of energy.
\end{itemize}

To the best knowledge of the author, such a presentation has still been missing from the literature.
However, the present work goes beyond merely reproducing already-known results.
In particular, we presented new inequalities for arbitrary driven processes, which also shed light on the more subtle distinction between several idealized variants.

Before we conclude, some additional comments on the conceptional basis of thermodynamics as a \emph{meta-}physical theory seem appropriate.

\subsubsection{On energy, time and probability}
At first glance, the generalized equilibrium distributions 
\begin{align}
  \den\rf(\vec{x}) \propto \exp(-\beta(E(\vec{x})-\ip\vec{\obs}(\vec{x})))  
  \label{eq:distribution-without-pars}
\end{align}
exhibit an apparent symmetry between the system energy $E(\vec{x})$ and phase-space functions $\vec{\obs}(\vec{x})$.
Indeed, for any non-driven process, the conjugate pairs $(\beta, E)$ and $(-\beta\gamma_i, \obs_i)$ of intensive and extensive quantities are mathematically equivalent.
This symmetry is then broken by the definition of work and heat via the time derivative of the expectation value $\eav{E} = \U$.
Hence, by looking only at Eq.~\eqref{eq:distribution-without-pars} a natural question comes to mind:
What makes energy special?

The answer to this question must necessarily come from physics, and not from mathematics.
In fact, it should arise from a \emph{fundamental} physical theory that captures universal physical laws on a sufficiently low level.
\emph{Energy is special because energy and dynamics, and thus the notion of time, are fundamentally related.}

Fundamental here means on a sufficiently low level, say Hamiltonian or quantum mechanics.
In these low-level descriptions, energy is characterized by the Hamiltonian which is responsible for the time evolution.
More abstractly, Noether's theorem shows that energy is the conserved quantity associated with the symmetry of physical laws with respect to time translations.
In the microcanonical derivation of statistical physics (\cf section~\ref{sec:canonical-ensemble}), the importance of the Hamiltonian (and thus energy) is evident from the start:
one \emph{defines} the entropy as the central quantity of statistical physics via the (logarithm of the) phase space volume $\W(\U)$ of all (physically distinct) microstates that are compatible with a given value $\U$ of the system's energy.

In contrast, the MaxEnt approach to statistical mechanics directly yields Eq.~\eqref{eq:distribution-without-pars} without the need to refer to a microscopic Hamiltonian.
Even though the connection between energy and entropy is more obscure, it still can be recovered.
Recall that uncertainty enters the microcanonical derivation in the form of equal \textit{a-priori} probabilities for all microscopic states on an energy shell.
Mathematically, this means that the volume of the energy shell is calculated using the natural normalized (Lebesgue or Liouville) measure.
The same assumption enters the MaxEnt approach, when we calculate the entropy of a probability density $\den(\vec{x})$:
The definition of probability densities already requires a natural reference measure, which in the present case is the natural Lebesgue measure.
Consequently, we calculate all integrals using the natural phase space volume $\df \vec{x}$.
Hence, using MaxEnt for (natural) phase space densities is as fundamental (or, as arbitrary) as starting the microcanonical derivation from the assumption of equal \textit{a-priori} probabilities.

However, this seems to be just another step down the rabbit hole.
One may ask for the origin of the natural measure on phase space.
Couldn't we just equivalently use another measure?
Coming back to Noether's theorem connecting energy and time we might also ask:
\emph{How does the natural measure (and thus probability) connect to the notion of time?}

There is also a consistent simple answer to that fundamental question:
\emph{The concept of time is nothing else than a fair weighting of dynamically accessible microstates in an inertial frame of reference.}
Recall that in a Newtonian inertial frame of reference a body remains at rest or moves with constant speed unless forces are present.
Hence, for a force-free particle, the notion of time simply expresses a uniform (and thus fair or unbiased) way of splitting the phase space states that are occupied by the trajectory of the particle.

\subsubsection{On reversibility and the arrow of time}
Another feature of fundamental evolution rules is their reversibility.
Leaving the weak force aside, on such a microscopic level, we are not able to infer the direction of time from the dynamics of the elementary constituents.
Moreover, the operation that reverses the microscopic arrow of time is a measure-preserving involution on the microscopic phase space and thus conserves entropy.
Notice that in Hamiltonian mechanics, the conservation of entropy is ensured by Liouville's theorem.

Relaxing irreversible dynamics are always coarse-grained descriptions of reality formulated on a certain level that describes an observer's phenomenology.
Like macroscopic entropy production, the thermodynamic arrow of time is an epistemic concept.
In the author's opinion, these connections between space and time,  probability and information might provide an interesting perspective on recent advances in cosmology~\cite{Bousso2002,Bekenstein2003,Hawking.etal2016,Hawking2014}.
Taking the universal discreteness of quantum mechanics (and the more fundamental CPT invariance) into account, this might even provide novel perspectives on quantum gravity.
However, at this point we leave the rabbit hole in order to conclude the present work.

\subsubsection{Outlook and conclusion}
\label{sec:conclusion}

In this work, we presented a first-principles approach to non-equilibrium thermodynamics based on the assumption of relaxing dynamics.
Such dynamics are forgetful or strictly non-anticipating, meaning that the distribution $\den\tind{\tau}$ of microscopic degrees of freedom gets less certain over time.
More precisely, ``less certain'' here means ``closer to the generalized equilibrium distribution $\den_\pars\rf$ for a physical system described by a set of thermodynamic parameters $\pars$''.
Acknowledging the fundamental meaning of energy and (microscopic) entropy, such distributions can similarly be obtained as the maximally non-committal distributions for a given set of phenomenological first-order constraints on extensive thermodynamic quantities.

Within our framework, we consistently identified general notions of a natural (generating) potential and the associated natural heat content, which generalizes the notion of enthalpy.
The process inequalities relating work, heat and information (like the Second Law, the Law of exergy degradation or Landauer's principle) were shown to be equivalent to the assumption of relaxing dynamics, which can be rationalized in an information-theoretic framework.

A careful distinction between driving and autonomous power emphasizes the importance of working with the correct (non-equilibrium) potentials if non-equilibrium inequalities are to be applied beyond the canonical, isothermal setup.
We have introduced a general bound for the irreversible work and discussed it for several real and idealized processes.
Further and maybe most importantly, we emphasized the importance of energy and entropy as the basic concepts of thermodynamics and discussed how they are intimately connected to dynamics and probability.

We have purposefully excluded the treatment of non-equilibrium steady states, because this requires are more detailed discussion of the notion of local equilibrium, which is beyond the scope of this paper.
For the same reason, we have not included a formulation of thermodynamics on the level of individual trajectories realized by a (Markovian) stochastic process.
A more detailed treatment of so-called Stochastic Thermodynamics (\cf Refs.~\cite{Ciliberto_etal2010,Seifert2012,VandenBroeck.Esposito2015} for an introduction), will be the subject of a future publication.
We believe that the applications of the present framework to so called fluctuation relations, which arise from comparing the probabilities forward and backward trajectories~\cite{Maes2004,Seifert2005}, can bring new insights.
In particular, studying the fluctuation relations for the process quantities introduced here (like driving, autonomous and mediate work as well as the gradient entropy flow) seems promising.

In conclusion, we see this work as a contribution towards a modern epistemic view of nonequilibrium thermodynamics, \ie a view of thermodynamics as a general physical theory of statistical inference.
It relates macroscopic (observable and controllable) and microscopic (hidden and uncontrollable) degrees of freedom.
This is very much in line of atomistic notion of heat that emerged around 1900.
After all, heat is an epistemic concept in the first place:
the energy of a system attributed to \emph{operationally inaccessible} degrees of freedom.

\vspace{1cm}
\acknowledgments
The author thanks the editors for the invitation to contribute to this special issue.
This work would not have been possible without insightful and sometimes fierce discussions with many people throughout the last years.
In particular, we acknowledge exchanges with
Lukas Geyrhofer,
Massimiliano Esposito,
David Hofmann,
Alexandre Lazarescu,
Matteo Polettini,
Riccardo Rao,
Hugo Touchette,
J\"urgen Vollmer
and
Artur Wachtel.
Various ThinkCamps hosted by the MPI-DS in G\"ottingen and the Institute of Entropology reassured the author to pursue these more conceptual issues.
The author is further indebted to his host institutions which enabled and encouraged a freedom of scientific thoughts, even if these thoughts were at times slightly tangential to project descriptions formulated \emph{a priori}.
This research was supported by the National Research Fund Luxembourg in the frame of the project FNR/A11/02.

\bibliography{relaxing}

\begin{thebibliography}{48}%
\makeatletter
\providecommand \@ifxundefined [1]{%
 \@ifx{#1\undefined}
}%
\providecommand \@ifnum [1]{%
 \ifnum #1\expandafter \@firstoftwo
 \else \expandafter \@secondoftwo
 \fi
}%
\providecommand \@ifx [1]{%
 \ifx #1\expandafter \@firstoftwo
 \else \expandafter \@secondoftwo
 \fi
}%
\providecommand \natexlab [1]{#1}%
\providecommand \enquote  [1]{``#1''}%
\providecommand \bibnamefont  [1]{#1}%
\providecommand \bibfnamefont [1]{#1}%
\providecommand \citenamefont [1]{#1}%
\providecommand \href@noop [0]{\@secondoftwo}%
\providecommand \href [0]{\begingroup \@sanitize@url \@href}%
\providecommand \@href[1]{\@@startlink{#1}\@@href}%
\providecommand \@@href[1]{\endgroup#1\@@endlink}%
\providecommand \@sanitize@url [0]{\catcode `\\12\catcode `\$12\catcode
  `\&12\catcode `\#12\catcode `\^12\catcode `\_12\catcode `\%12\relax}%
\providecommand \@@startlink[1]{}%
\providecommand \@@endlink[0]{}%
\providecommand \url  [0]{\begingroup\@sanitize@url \@url }%
\providecommand \@url [1]{\endgroup\@href {#1}{\urlprefix }}%
\providecommand \urlprefix  [0]{URL }%
\providecommand \Eprint [0]{\href }%
\providecommand \doibase [0]{http://dx.doi.org/}%
\providecommand \selectlanguage [0]{\@gobble}%
\providecommand \bibinfo  [0]{\@secondoftwo}%
\providecommand \bibfield  [0]{\@secondoftwo}%
\providecommand \translation [1]{[#1]}%
\providecommand \BibitemOpen [0]{}%
\providecommand \bibitemStop [0]{}%
\providecommand \bibitemNoStop [0]{.\EOS\space}%
\providecommand \EOS [0]{\spacefactor3000\relax}%
\providecommand \BibitemShut  [1]{\csname bibitem#1\endcsname}%
\let\auto@bib@innerbib\@empty
\bibitem [{\citenamefont {Peliti}\ and\ \citenamefont
  {Rechtman}(2017)}]{Peliti.Rechtman2017}%
  \BibitemOpen
  \bibfield  {author} {\bibinfo {author} {\bibfnamefont {L.}~\bibnamefont
  {Peliti}}\ and\ \bibinfo {author} {\bibfnamefont {R.}~\bibnamefont
  {Rechtman}},\ }\href {\doibase 10.1007/s10955-016-1615-8} {\bibfield
  {journal} {\bibinfo  {journal} {Journal of Statistical Physics}\ }\textbf
  {\bibinfo {volume} {167}},\ \bibinfo {pages} {1020} (\bibinfo {year}
  {2017})}\BibitemShut {NoStop}%
\bibitem [{\citenamefont {Shannon}(1948)}]{Shannon1948}%
  \BibitemOpen
  \bibfield  {author} {\bibinfo {author} {\bibfnamefont {C.~E.}\ \bibnamefont
  {Shannon}},\ }\href@noop {} {\bibfield  {journal} {\bibinfo  {journal} {AT\&T
  Technical Journal}\ }\textbf {\bibinfo {volume} {27}},\ \bibinfo {pages}
  {379} (\bibinfo {year} {1948})},\ \bibinfo {note} {many reprints
  available}\BibitemShut {NoStop}%
\bibitem [{\citenamefont {Jaynes}(1957)}]{Jaynes1957}%
  \BibitemOpen
  \bibfield  {author} {\bibinfo {author} {\bibfnamefont {E.~T.}\ \bibnamefont
  {Jaynes}},\ }\href {\doibase 10.1103/PhysRev.106.620} {\bibfield  {journal}
  {\bibinfo  {journal} {Physical Review}\ }\textbf {\bibinfo {volume} {106}},\
  \bibinfo {pages} {620} (\bibinfo {year} {1957})}\BibitemShut {NoStop}%
\bibitem [{\citenamefont {Landauer}(1961)}]{Landauer1961}%
  \BibitemOpen
  \bibfield  {author} {\bibinfo {author} {\bibfnamefont {R.}~\bibnamefont
  {Landauer}},\ }\href@noop {} {\bibfield  {journal} {\bibinfo  {journal} {IBM
  Journal of Research and Development}\ }\textbf {\bibinfo {volume} {5}},\
  \bibinfo {pages} {183} (\bibinfo {year} {1961})}\BibitemShut {NoStop}%
\bibitem [{\citenamefont {Procaccia}\ and\ \citenamefont
  {Levine}(1976)}]{Procaccia.Levine1976}%
  \BibitemOpen
  \bibfield  {author} {\bibinfo {author} {\bibfnamefont {I.}~\bibnamefont
  {Procaccia}}\ and\ \bibinfo {author} {\bibfnamefont {R.}~\bibnamefont
  {Levine}},\ }\href@noop {} {\bibfield  {journal} {\bibinfo  {journal} {The
  Journal of Chemical Physics}\ }\textbf {\bibinfo {volume} {65}},\ \bibinfo
  {pages} {3357} (\bibinfo {year} {1976})}\BibitemShut {NoStop}%
\bibitem [{\citenamefont {Bennett}(1982)}]{Bennett1982}%
  \BibitemOpen
  \bibfield  {author} {\bibinfo {author} {\bibfnamefont {C.~H.}\ \bibnamefont
  {Bennett}},\ }\href@noop {} {\bibfield  {journal} {\bibinfo  {journal}
  {International Journal of Theoretical Physics}\ }\textbf {\bibinfo {volume}
  {21}},\ \bibinfo {pages} {905} (\bibinfo {year} {1982})}\BibitemShut
  {NoStop}%
\bibitem [{\citenamefont {Toyabe}\ \emph {et~al.}(2010)\citenamefont {Toyabe},
  \citenamefont {Sagawa}, \citenamefont {Ueda}, \citenamefont {Muneyuki},\ and\
  \citenamefont {Sano}}]{Toyabe_etal2010}%
  \BibitemOpen
  \bibfield  {author} {\bibinfo {author} {\bibfnamefont {S.}~\bibnamefont
  {Toyabe}}, \bibinfo {author} {\bibfnamefont {T.}~\bibnamefont {Sagawa}},
  \bibinfo {author} {\bibfnamefont {M.}~\bibnamefont {Ueda}}, \bibinfo {author}
  {\bibfnamefont {E.}~\bibnamefont {Muneyuki}}, \ and\ \bibinfo {author}
  {\bibfnamefont {M.}~\bibnamefont {Sano}},\ }\href@noop {} {\bibfield
  {journal} {\bibinfo  {journal} {Nat. Phys.}\ }\textbf {\bibinfo {volume}
  {6}},\ \bibinfo {pages} {988} (\bibinfo {year} {2010})}\BibitemShut {NoStop}%
\bibitem [{\citenamefont {Esposito}\ and\ \citenamefont {Van~den
  Broeck}(2011)}]{Esposito.VandenBroeck2011}%
  \BibitemOpen
  \bibfield  {author} {\bibinfo {author} {\bibfnamefont {M.}~\bibnamefont
  {Esposito}}\ and\ \bibinfo {author} {\bibfnamefont {C.}~\bibnamefont {Van~den
  Broeck}},\ }\href {http://stacks.iop.org/0295-5075/95/i=4/a=40004} {\bibfield
   {journal} {\bibinfo  {journal} {EPL (Europhysics Letters)}\ }\textbf
  {\bibinfo {volume} {95}},\ \bibinfo {pages} {40004} (\bibinfo {year}
  {2011})}\BibitemShut {NoStop}%
\bibitem [{\citenamefont {Sagawa}(2012)}]{Sagawa2012}%
  \BibitemOpen
  \bibfield  {author} {\bibinfo {author} {\bibfnamefont {T.}~\bibnamefont
  {Sagawa}},\ }\href@noop {} {\emph {\bibinfo {title} {Thermodynamics of
  information processing in small systems}}}\ (\bibinfo  {publisher} {Springer
  Verlag},\ \bibinfo {year} {2012})\BibitemShut {NoStop}%
\bibitem [{\citenamefont {B{\'e}rut}\ \emph {et~al.}(2012)\citenamefont
  {B{\'e}rut}, \citenamefont {Arakelyan}, \citenamefont {Petrosyan},
  \citenamefont {Ciliberto}, \citenamefont {Dillenschneider},\ and\
  \citenamefont {Lutz}}]{Berut_etal2012}%
  \BibitemOpen
  \bibfield  {author} {\bibinfo {author} {\bibfnamefont {A.}~\bibnamefont
  {B{\'e}rut}}, \bibinfo {author} {\bibfnamefont {A.}~\bibnamefont
  {Arakelyan}}, \bibinfo {author} {\bibfnamefont {A.}~\bibnamefont
  {Petrosyan}}, \bibinfo {author} {\bibfnamefont {S.}~\bibnamefont
  {Ciliberto}}, \bibinfo {author} {\bibfnamefont {R.}~\bibnamefont
  {Dillenschneider}}, \ and\ \bibinfo {author} {\bibfnamefont {E.}~\bibnamefont
  {Lutz}},\ }\href@noop {} {\bibfield  {journal} {\bibinfo  {journal} {Nature}\
  }\textbf {\bibinfo {volume} {483}},\ \bibinfo {pages} {187} (\bibinfo {year}
  {2012})}\BibitemShut {NoStop}%
\bibitem [{\citenamefont {Horowitz}\ \emph {et~al.}(2013)\citenamefont
  {Horowitz}, \citenamefont {Sagawa},\ and\ \citenamefont
  {Parrondo}}]{Horowitz_etal2013}%
  \BibitemOpen
  \bibfield  {author} {\bibinfo {author} {\bibfnamefont {J.~M.}\ \bibnamefont
  {Horowitz}}, \bibinfo {author} {\bibfnamefont {T.}~\bibnamefont {Sagawa}}, \
  and\ \bibinfo {author} {\bibfnamefont {J.~M.}\ \bibnamefont {Parrondo}},\
  }\href@noop {} {\bibfield  {journal} {\bibinfo  {journal} {Physical review
  letters}\ }\textbf {\bibinfo {volume} {111}},\ \bibinfo {pages} {010602}
  (\bibinfo {year} {2013})}\BibitemShut {NoStop}%
\bibitem [{\citenamefont {Barato}\ \emph {et~al.}(2014)\citenamefont {Barato},
  \citenamefont {Hartich},\ and\ \citenamefont {Seifert}}]{Barato.etal2014}%
  \BibitemOpen
  \bibfield  {author} {\bibinfo {author} {\bibfnamefont {A.~C.}\ \bibnamefont
  {Barato}}, \bibinfo {author} {\bibfnamefont {D.}~\bibnamefont {Hartich}}, \
  and\ \bibinfo {author} {\bibfnamefont {U.}~\bibnamefont {Seifert}},\ }\href
  {\doibase 10.1088/1367-2630/16/10/103024} {\bibfield  {journal} {\bibinfo
  {journal} {New J. Phys.}\ }\textbf {\bibinfo {volume} {16}},\ \bibinfo
  {pages} {103024} (\bibinfo {year} {2014})}\BibitemShut {NoStop}%
\bibitem [{\citenamefont {Parrondo}\ \emph {et~al.}(2015)\citenamefont
  {Parrondo}, \citenamefont {Horowitz},\ and\ \citenamefont
  {Sagawa}}]{Parrondo.etal2015}%
  \BibitemOpen
  \bibfield  {author} {\bibinfo {author} {\bibfnamefont {J.~M.~R.}\
  \bibnamefont {Parrondo}}, \bibinfo {author} {\bibfnamefont {J.~M.}\
  \bibnamefont {Horowitz}}, \ and\ \bibinfo {author} {\bibfnamefont
  {T.}~\bibnamefont {Sagawa}},\ }\href {\doibase 10.1038/nphys3230} {\bibfield
  {journal} {\bibinfo  {journal} {Nat. Phys.}\ }\textbf {\bibinfo {volume}
  {11}},\ \bibinfo {pages} {131} (\bibinfo {year} {2015})}\BibitemShut
  {NoStop}%
\bibitem [{\citenamefont {Sartori}\ and\ \citenamefont
  {Pigolotti}(2015)}]{Sartori.Pigolotti2015}%
  \BibitemOpen
  \bibfield  {author} {\bibinfo {author} {\bibfnamefont {P.}~\bibnamefont
  {Sartori}}\ and\ \bibinfo {author} {\bibfnamefont {S.}~\bibnamefont
  {Pigolotti}},\ }\href {\doibase 10.1103/PhysRevX.5.041039} {\bibfield
  {journal} {\bibinfo  {journal} {Phys. Rev. X}\ }\textbf {\bibinfo {volume}
  {5}},\ \bibinfo {pages} {041039} (\bibinfo {year} {2015})}\BibitemShut
  {NoStop}%
\bibitem [{\citenamefont {Rao}\ and\ \citenamefont
  {Esposito}(2016)}]{Rao.Esposito2016}%
  \BibitemOpen
  \bibfield  {author} {\bibinfo {author} {\bibfnamefont {R.}~\bibnamefont
  {Rao}}\ and\ \bibinfo {author} {\bibfnamefont {M.}~\bibnamefont {Esposito}},\
  }\href {\doibase 10.1103/PhysRevX.6.041064} {\bibfield  {journal} {\bibinfo
  {journal} {Phys. Rev. X}\ }\textbf {\bibinfo {volume} {6}},\ \bibinfo {pages}
  {041064} (\bibinfo {year} {2016})}\BibitemShut {NoStop}%
\bibitem [{\citenamefont {Schilpp}\ and\ \citenamefont
  {Jehle}(1951)}]{Schilpp.Jehle1951}%
  \BibitemOpen
  \bibfield  {author} {\bibinfo {author} {\bibfnamefont {P.~A.}\ \bibnamefont
  {Schilpp}}\ and\ \bibinfo {author} {\bibfnamefont {H.}~\bibnamefont
  {Jehle}},\ }\href@noop {} {\bibfield  {journal} {\bibinfo  {journal}
  {American Journal of Physics}\ }\textbf {\bibinfo {volume} {19}},\ \bibinfo
  {pages} {252} (\bibinfo {year} {1951})}\BibitemShut {NoStop}%
\bibitem [{\citenamefont {Jaynes}(1965)}]{Jaynes1965}%
  \BibitemOpen
  \bibfield  {author} {\bibinfo {author} {\bibfnamefont {E.~T.}\ \bibnamefont
  {Jaynes}},\ }\href {\doibase 10.1119/1.1971557} {\bibfield  {journal}
  {\bibinfo  {journal} {American Journal of Physics}\ }\textbf {\bibinfo
  {volume} {33}},\ \bibinfo {pages} {391} (\bibinfo {year} {1965})}\BibitemShut
  {NoStop}%
\bibitem [{\citenamefont {Shore}\ and\ \citenamefont
  {Johnson}(1980)}]{Shore.Johnson1980}%
  \BibitemOpen
  \bibfield  {author} {\bibinfo {author} {\bibfnamefont {J.}~\bibnamefont
  {Shore}}\ and\ \bibinfo {author} {\bibfnamefont {R.}~\bibnamefont
  {Johnson}},\ }\href {\doibase 10.1109/TIT.1980.1056144} {\bibfield  {journal}
  {\bibinfo  {journal} {IEEE Transactions on Information Theory}\ }\textbf
  {\bibinfo {volume} {26}},\ \bibinfo {pages} {26} (\bibinfo {year}
  {1980})}\BibitemShut {NoStop}%
\bibitem [{\citenamefont {Jarzynski}(1997)}]{Jarzynski1997}%
  \BibitemOpen
  \bibfield  {author} {\bibinfo {author} {\bibfnamefont {C.}~\bibnamefont
  {Jarzynski}},\ }\href@noop {} {\bibfield  {journal} {\bibinfo  {journal}
  {Phys. Rev. Lett.}\ }\textbf {\bibinfo {volume} {78}},\ \bibinfo {pages}
  {2690} (\bibinfo {year} {1997})}\BibitemShut {NoStop}%
\bibitem [{\citenamefont {Vaikuntanathan}\ and\ \citenamefont
  {Jarzynski}(2009)}]{Vaikuntanathan.Jarzynski2009}%
  \BibitemOpen
  \bibfield  {author} {\bibinfo {author} {\bibfnamefont {S.}~\bibnamefont
  {Vaikuntanathan}}\ and\ \bibinfo {author} {\bibfnamefont {C.}~\bibnamefont
  {Jarzynski}},\ }\href@noop {} {\bibfield  {journal} {\bibinfo  {journal} {EPL
  (Europhysics Letters)}\ }\textbf {\bibinfo {volume} {87}},\ \bibinfo {pages}
  {60005} (\bibinfo {year} {2009})}\BibitemShut {NoStop}%
\bibitem [{\citenamefont {Takara}\ \emph {et~al.}(2010)\citenamefont {Takara},
  \citenamefont {Hasegawa},\ and\ \citenamefont {Driebe}}]{Takara.etal2010}%
  \BibitemOpen
  \bibfield  {author} {\bibinfo {author} {\bibfnamefont {K.}~\bibnamefont
  {Takara}}, \bibinfo {author} {\bibfnamefont {H.-H.}\ \bibnamefont
  {Hasegawa}}, \ and\ \bibinfo {author} {\bibfnamefont {D.}~\bibnamefont
  {Driebe}},\ }\href@noop {} {\bibfield  {journal} {\bibinfo  {journal}
  {Physics Letters A}\ }\textbf {\bibinfo {volume} {375}},\ \bibinfo {pages}
  {88} (\bibinfo {year} {2010})}\BibitemShut {NoStop}%
\bibitem [{\citenamefont {Gibbs}(1902)}]{Gibbs1902}%
  \BibitemOpen
  \bibfield  {author} {\bibinfo {author} {\bibfnamefont {J.~W.}\ \bibnamefont
  {Gibbs}},\ }\href@noop {} {\emph {\bibinfo {title} {Elementary {Principles}
  in {Statistical} {Mechanics}: {Developed} with {Especial} {Reference} to the
  {Rational} {Foundations} of {Thermodynamics}}}}\ (\bibinfo  {publisher} {C.
  Scribner's sons},\ \bibinfo {address} {New York},\ \bibinfo {year}
  {1902})\BibitemShut {NoStop}%
\bibitem [{\citenamefont {Callen}(1998)}]{Callen1998}%
  \BibitemOpen
  \bibfield  {author} {\bibinfo {author} {\bibfnamefont {H.~B.}\ \bibnamefont
  {Callen}},\ }\href@noop {} {\emph {\bibinfo {title} {Thermodynamics and an
  Introduction to Thermostatistics}}}\ (\bibinfo  {publisher} {AAPT},\ \bibinfo
  {year} {1998})\BibitemShut {NoStop}%
\bibitem [{\citenamefont {Penrose}(1970)}]{Penrose1970}%
  \BibitemOpen
  \bibfield  {author} {\bibinfo {author} {\bibfnamefont {O.}~\bibnamefont
  {Penrose}},\ }\href@noop {} {\emph {\bibinfo {title} {{Foundations of
  statistical mechanics; a deductive treatment}}}}\ (\bibinfo  {publisher}
  {Pergamon Press},\ \bibinfo {address} {Oxford},\ \bibinfo {year}
  {1970})\BibitemShut {NoStop}%
\bibitem [{\citenamefont {Touchette}(2009)}]{Touchette2009}%
  \BibitemOpen
  \bibfield  {author} {\bibinfo {author} {\bibfnamefont {H.}~\bibnamefont
  {Touchette}},\ }\href {\doibase 10.1016/j.physrep.2009.05.002} {\bibfield
  {journal} {\bibinfo  {journal} {Phys. Rep.}\ }\textbf {\bibinfo {volume}
  {478}},\ \bibinfo {pages} {1} (\bibinfo {year} {2009})}\BibitemShut {NoStop}%
\bibitem [{\citenamefont {Balian}(2007)}]{Balian2007}%
  \BibitemOpen
  \bibfield  {author} {\bibinfo {author} {\bibfnamefont {R.}~\bibnamefont
  {Balian}},\ }\href@noop {} {\emph {\bibinfo {title} {From microphysics to
  macrophysics: methods and applications of statistical physics}}},\
  Vol.~\bibinfo {volume} {2}\ (\bibinfo  {publisher} {Springer Science \&
  Business Media},\ \bibinfo {year} {2007})\BibitemShut {NoStop}%
\bibitem [{\citenamefont {Ellis}(2012)}]{Ellis2012}%
  \BibitemOpen
  \bibfield  {author} {\bibinfo {author} {\bibfnamefont {R.~S.}\ \bibnamefont
  {Ellis}},\ }\href@noop {} {\emph {\bibinfo {title} {Entropy, large
  deviations, and statistical mechanics}}},\ Vol.\ \bibinfo {volume} {271}\
  (\bibinfo  {publisher} {Springer Science \& Business Media},\ \bibinfo {year}
  {2012})\BibitemShut {NoStop}%
\bibitem [{\citenamefont {Chetrite}\ and\ \citenamefont
  {Touchette}(2014)}]{Chetrite.Touchette2014}%
  \BibitemOpen
  \bibfield  {author} {\bibinfo {author} {\bibfnamefont {R.}~\bibnamefont
  {Chetrite}}\ and\ \bibinfo {author} {\bibfnamefont {H.}~\bibnamefont
  {Touchette}},\ }\href {\doibase 10.1007/s00023-014-0375-8} {\bibfield
  {journal} {\bibinfo  {journal} {Annales Henri Poincaré}\ }\textbf {\bibinfo
  {volume} {16}},\ \bibinfo {pages} {2005} (\bibinfo {year}
  {2014})}\BibitemShut {NoStop}%
\bibitem [{\citenamefont {Cover}\ and\ \citenamefont
  {Thomas}(2006)}]{Cover.Thomas2006}%
  \BibitemOpen
  \bibfield  {author} {\bibinfo {author} {\bibfnamefont {T.~M.}\ \bibnamefont
  {Cover}}\ and\ \bibinfo {author} {\bibfnamefont {J.~A.}\ \bibnamefont
  {Thomas}},\ }\href@noop {} {\emph {\bibinfo {title} {Elements of Information
  Theory (Wiley Series in Telecommunications and Signal Processing)}}}\
  (\bibinfo  {publisher} {Wiley-Interscience},\ \bibinfo {year}
  {2006})\BibitemShut {NoStop}%
\bibitem [{\citenamefont {Schnakenberg}(1976)}]{Schnakenberg1976}%
  \BibitemOpen
  \bibfield  {author} {\bibinfo {author} {\bibfnamefont {J.}~\bibnamefont
  {Schnakenberg}},\ }\href {\doibase 10.1103/RevModPhys.48.571} {\bibfield
  {journal} {\bibinfo  {journal} {Rev. Mod. Phys.}\ }\textbf {\bibinfo {volume}
  {48}},\ \bibinfo {pages} {571} (\bibinfo {year} {1976})}\BibitemShut
  {NoStop}%
\bibitem [{\citenamefont {Baiesi}\ and\ \citenamefont
  {Maes}(2013)}]{Baiesi.Maes2013}%
  \BibitemOpen
  \bibfield  {author} {\bibinfo {author} {\bibfnamefont {M.}~\bibnamefont
  {Baiesi}}\ and\ \bibinfo {author} {\bibfnamefont {C.}~\bibnamefont {Maes}},\
  }\href {http://stacks.iop.org/1367-2630/15/i=1/a=013004} {\bibfield
  {journal} {\bibinfo  {journal} {New Journal of Physics}\ }\textbf {\bibinfo
  {volume} {15}},\ \bibinfo {pages} {013004} (\bibinfo {year}
  {2013})}\BibitemShut {NoStop}%
\bibitem [{\citenamefont {Wachtel}\ \emph {et~al.}(2015)\citenamefont
  {Wachtel}, \citenamefont {Vollmer},\ and\ \citenamefont
  {Altaner}}]{Wachtel.etal2015}%
  \BibitemOpen
  \bibfield  {author} {\bibinfo {author} {\bibfnamefont {A.}~\bibnamefont
  {Wachtel}}, \bibinfo {author} {\bibfnamefont {J.}~\bibnamefont {Vollmer}}, \
  and\ \bibinfo {author} {\bibfnamefont {B.}~\bibnamefont {Altaner}},\ }\href
  {\doibase 10.1103/physreve.92.042132} {\bibfield  {journal} {\bibinfo
  {journal} {Phys. Rev. E}\ }\textbf {\bibinfo {volume} {92}},\ \bibinfo
  {pages} {042132} (\bibinfo {year} {2015})}\BibitemShut {NoStop}%
\bibitem [{\citenamefont {Esposito}\ and\ \citenamefont {Van~den
  Broeck}(2010)}]{Esposito.VandenBroeck2010}%
  \BibitemOpen
  \bibfield  {author} {\bibinfo {author} {\bibfnamefont {M.}~\bibnamefont
  {Esposito}}\ and\ \bibinfo {author} {\bibfnamefont {C.}~\bibnamefont {Van~den
  Broeck}},\ }\href {\doibase 10.1103/physreve.82.011143} {\bibfield  {journal}
  {\bibinfo  {journal} {Phys. Rev. E}\ }\textbf {\bibinfo {volume} {82}},\
  \bibinfo {pages} {011143} (\bibinfo {year} {2010})}\BibitemShut {NoStop}%
\bibitem [{\citenamefont {Esposito}(2012)}]{Esposito2012}%
  \BibitemOpen
  \bibfield  {author} {\bibinfo {author} {\bibfnamefont {M.}~\bibnamefont
  {Esposito}},\ }\href {\doibase 10.1103/PhysRevE.85.041125} {\bibfield
  {journal} {\bibinfo  {journal} {Phys. Rev. E}\ }\textbf {\bibinfo {volume}
  {85}},\ \bibinfo {pages} {041125} (\bibinfo {year} {2012})}\BibitemShut
  {NoStop}%
\bibitem [{\citenamefont {Seifert}(2011)}]{Seifert2011}%
  \BibitemOpen
  \bibfield  {author} {\bibinfo {author} {\bibfnamefont {U.}~\bibnamefont
  {Seifert}},\ }\href {\doibase 10.1140/epje/i2011-11026-7} {\bibfield
  {journal} {\bibinfo  {journal} {Eur. Phys. J. E}\ }\textbf {\bibinfo {volume}
  {34}},\ \bibinfo {eid} {26} (\bibinfo {year} {2011})}\BibitemShut {NoStop}%
\bibitem [{\citenamefont {Markowich}\ and\ \citenamefont
  {Villani}(2000)}]{Markowich.Villani2000}%
  \BibitemOpen
  \bibfield  {author} {\bibinfo {author} {\bibfnamefont {P.~A.}\ \bibnamefont
  {Markowich}}\ and\ \bibinfo {author} {\bibfnamefont {C.}~\bibnamefont
  {Villani}},\ }\href@noop {} {\bibfield  {journal} {\bibinfo  {journal} {Mat.
  Contemp}\ }\textbf {\bibinfo {volume} {19}},\ \bibinfo {pages} {1} (\bibinfo
  {year} {2000})}\BibitemShut {NoStop}%
\bibitem [{\citenamefont {Lozano}\ and\ \citenamefont
  {Valero}(1993)}]{Lozano.Valero1993}%
  \BibitemOpen
  \bibfield  {author} {\bibinfo {author} {\bibfnamefont {M.}~\bibnamefont
  {Lozano}}\ and\ \bibinfo {author} {\bibfnamefont {A.}~\bibnamefont
  {Valero}},\ }\href@noop {} {\bibfield  {journal} {\bibinfo  {journal}
  {Energy}\ }\textbf {\bibinfo {volume} {18}},\ \bibinfo {pages} {939}
  (\bibinfo {year} {1993})}\BibitemShut {NoStop}%
\bibitem [{\citenamefont {Dincer}\ and\ \citenamefont
  {Rosen}(2012)}]{Dincer.Rosen2012}%
  \BibitemOpen
  \bibfield  {author} {\bibinfo {author} {\bibfnamefont {I.}~\bibnamefont
  {Dincer}}\ and\ \bibinfo {author} {\bibfnamefont {M.~A.}\ \bibnamefont
  {Rosen}},\ }\href@noop {} {\emph {\bibinfo {title} {Exergy: energy,
  environment and sustainable development}}}\ (\bibinfo  {publisher} {Newnes},\
  \bibinfo {year} {2012})\BibitemShut {NoStop}%
\bibitem [{\citenamefont {Schmiedl}\ and\ \citenamefont
  {Seifert}(2007)}]{Schmiedl.Seifert2007}%
  \BibitemOpen
  \bibfield  {author} {\bibinfo {author} {\bibfnamefont {T.}~\bibnamefont
  {Schmiedl}}\ and\ \bibinfo {author} {\bibfnamefont {U.}~\bibnamefont
  {Seifert}},\ }\href {\doibase 10.1063/1.2428297} {\bibfield  {journal}
  {\bibinfo  {journal} {J. Chem. Phys.}\ }\textbf {\bibinfo {volume} {126}},\
  \bibinfo {pages} {044101} (\bibinfo {year} {2007})}\BibitemShut {NoStop}%
\bibitem [{\citenamefont {Bousso}(2002)}]{Bousso2002}%
  \BibitemOpen
  \bibfield  {author} {\bibinfo {author} {\bibfnamefont {R.}~\bibnamefont
  {Bousso}},\ }\href@noop {} {\bibfield  {journal} {\bibinfo  {journal} {Rev.
  Mod. Phys.}\ }\textbf {\bibinfo {volume} {74}},\ \bibinfo {pages} {825}
  (\bibinfo {year} {2002})}\BibitemShut {NoStop}%
\bibitem [{\citenamefont {Bekenstein}(2003)}]{Bekenstein2003}%
  \BibitemOpen
  \bibfield  {author} {\bibinfo {author} {\bibfnamefont {J.}~\bibnamefont
  {Bekenstein}},\ }\href@noop {} {\bibfield  {journal} {\bibinfo  {journal}
  {Sci. Am.}\ }\textbf {\bibinfo {volume} {289}},\ \bibinfo {pages} {58}
  (\bibinfo {year} {2003})}\BibitemShut {NoStop}%
\bibitem [{\citenamefont {Hawking}\ \emph {et~al.}(2016)\citenamefont
  {Hawking}, \citenamefont {Perry},\ and\ \citenamefont
  {Strominger}}]{Hawking.etal2016}%
  \BibitemOpen
  \bibfield  {author} {\bibinfo {author} {\bibfnamefont {S.~W.}\ \bibnamefont
  {Hawking}}, \bibinfo {author} {\bibfnamefont {M.~J.}\ \bibnamefont {Perry}},
  \ and\ \bibinfo {author} {\bibfnamefont {A.}~\bibnamefont {Strominger}},\
  }\href {\doibase 10.1103/PhysRevLett.116.231301} {\bibfield  {journal}
  {\bibinfo  {journal} {Phys. Rev. Lett.}\ }\textbf {\bibinfo {volume} {116}},\
  \bibinfo {pages} {231301} (\bibinfo {year} {2016})}\BibitemShut {NoStop}%
\bibitem [{\citenamefont {Hawking}(2014)}]{Hawking2014}%
  \BibitemOpen
  \bibfield  {author} {\bibinfo {author} {\bibfnamefont {S.~W.}\ \bibnamefont
  {Hawking}},\ }\href@noop {} {\bibfield  {journal} {\bibinfo  {journal} {arXiv
  preprint arXiv:1401.5761}\ } (\bibinfo {year} {2014})}\BibitemShut {NoStop}%
\bibitem [{\citenamefont {Ciliberto}\ \emph {et~al.}(2010)\citenamefont
  {Ciliberto}, \citenamefont {Joubaud},\ and\ \citenamefont
  {Petrosyan}}]{Ciliberto_etal2010}%
  \BibitemOpen
  \bibfield  {author} {\bibinfo {author} {\bibfnamefont {S.}~\bibnamefont
  {Ciliberto}}, \bibinfo {author} {\bibfnamefont {S.}~\bibnamefont {Joubaud}},
  \ and\ \bibinfo {author} {\bibfnamefont {A.}~\bibnamefont {Petrosyan}},\
  }\href@noop {} {\bibfield  {journal} {\bibinfo  {journal} {J. Stat. Mech.}\
  }\textbf {\bibinfo {volume} {2010}},\ \bibinfo {pages} {P12003} (\bibinfo
  {year} {2010})}\BibitemShut {NoStop}%
\bibitem [{\citenamefont {Seifert}(2012)}]{Seifert2012}%
  \BibitemOpen
  \bibfield  {author} {\bibinfo {author} {\bibfnamefont {U.}~\bibnamefont
  {Seifert}},\ }\href {\doibase 10.1088/0034-4885/75/12/126001} {\bibfield
  {journal} {\bibinfo  {journal} {Rep. Prog. Phys.}\ }\textbf {\bibinfo
  {volume} {75}},\ \bibinfo {pages} {126001} (\bibinfo {year}
  {2012})}\BibitemShut {NoStop}%
\bibitem [{\citenamefont {Van~den Broeck}\ and\ \citenamefont
  {Esposito}(2015)}]{VandenBroeck.Esposito2015}%
  \BibitemOpen
  \bibfield  {author} {\bibinfo {author} {\bibfnamefont {C.}~\bibnamefont
  {Van~den Broeck}}\ and\ \bibinfo {author} {\bibfnamefont {M.}~\bibnamefont
  {Esposito}},\ }\href {\doibase 10.1016/j.physa.2014.04.035} {\bibfield
  {journal} {\bibinfo  {journal} {Physica A}\ }\textbf {\bibinfo {volume}
  {418}},\ \bibinfo {pages} {6} (\bibinfo {year} {2015})},\ \bibinfo {note}
  {proceedings of the 13th International Summer School on Fundamental Problems
  in Statistical Physics}\BibitemShut {NoStop}%
\bibitem [{\citenamefont {Maes}(2004)}]{Maes2004}%
  \BibitemOpen
  \bibfield  {author} {\bibinfo {author} {\bibfnamefont {C.}~\bibnamefont
  {Maes}},\ }in\ \href@noop {} {\emph {\bibinfo {booktitle} {Poincar{\'e}
  Seminar 2003: Bose-Einstein condensation-entropy}}}\ (\bibinfo  {publisher}
  {Birkh{\"a}user, Basel},\ \bibinfo {year} {2004})\ p.\ \bibinfo {pages}
  {145}\BibitemShut {NoStop}%
\bibitem [{\citenamefont {Seifert}(2005)}]{Seifert2005}%
  \BibitemOpen
  \bibfield  {author} {\bibinfo {author} {\bibfnamefont {U.}~\bibnamefont
  {Seifert}},\ }\href@noop {} {\bibfield  {journal} {\bibinfo  {journal} {Phys.
  Rev. Lett.}\ }\textbf {\bibinfo {volume} {95}},\ \bibinfo {pages} {40602}
  (\bibinfo {year} {2005})}\BibitemShut {NoStop}%
\end{thebibliography}%

\end{document}